\documentclass[fleqn,usenatbib]{mnras}

\usepackage{newtxtext,newtxmath}

\usepackage[T1]{fontenc}
\usepackage{ae,aecompl}

\usepackage{graphicx}	
\usepackage{amsmath}	

\title[Non-solar abundance ratios trends of dEs]{Non-solar abundance ratios trends of dEs in Fornax Cluster using newly defined high resolution indices}

\author[\c{S}. \c{S}en et al.]{
\c{S}eyda \c{S}en$^{1,2}$\thanks{E-mail: senseydastar@gmail.com},
Reynier F. Peletier$^{1}$, Alexandre Vazdekis$^{3,4}$
\\
\\
$^{1}$Kapteyn Astronomical Institute, University of Groningen, P. O. Box 800, 9700 AV Groningen, Netherlands\\
$^{2}$Sabancı University, Faculty of Engineering and Natural Sciences, İstanbul 34956, Turkey\\
$^{3}$ Instituto de Astrof{\'i}sica de Canarias,Calle V{\'i}a L{\'a}ctea s/n, E-38200 La Laguna, Tenerife, Spain\\
$^{4}$ Departamento de Astrof\'isica, Universidad de La Laguna, E-38205 La Laguna, Tenerife, Spain
}
\date{Accepted XXX. Received YYY; in original form ZZZ}

\pubyear{2022}

\begin{document}
\label{firstpage}
\pagerange{\pageref{firstpage}--\pageref{lastpage}}
\maketitle

\begin{abstract}
We perform a detailed study of  the stellar populations in a sample of massive Fornax dwarf galaxies using a set of newly defined line indices. Using data from the Integral field spectroscopic data, we study abundance ratios of eight dEs with stellar mass ranging from 10$^8$ to 10$^{9.5}$ M$_\odot$ in the Fornax cluster. We present the definitions of a new set of high-resolution Lick-style indices to be used for stellar population studies of unresolved small stellar systems.  We identify 23 absorption features and continuum regions, mainly dominated by 12 elements (Na, Ca, Sc, Ti, V, Cr, Mn, Fe, Ni, Y, Ba and Nd) in the wavelength range 4700 - 5400 \AA\ and characterise them as a function of age, metallicity and alpha element abundance ratios. We analyse eight dEs and interpret the line strengths, measured in our new high resolution system of indices, with the aid of stellar population models with high enough spectral resolution. We obtain abundance ratio proxies for a number of elements that have never been studied before for dwarf ellipticals outside the Local Group. These proxies represent relative deviations from predicted index-strengths of base stellar population models built-up following the abundance pattern of The Galaxy. The abundance proxy trend results are compared to abundance ratios from resolved stars in the Local Group, and indices from integrated light of larger early-type galaxies. We find that all our dwarfs show a pattern of abundance ratios consistent with the disk of the Milky Way, indicative of slow formation in comparison to their high mass counterparts. 
\end{abstract}




\section{Introduction}
The stellar populations of galaxies provide a fossil record of their formation and evolutionary history. An important tool needed to study galaxy evolution is stellar population synthesis. The stellar content and chemical composition of the unresolved stellar populations of galaxies can be obtained by detailed study of the observed absorption features present in their integrated spectra, using the information in the continuum and the absorption line strengths.

Dwarf elliptical galaxies (dEs) are known to exist in large numbers in galaxy clusters \citet{sandageandbinggeli.1984AJ.....89..919S}. They give us the opportunity to study the star formation history and chemical evolution not only of these galaxies, but also of galaxy clusters themselves (e.g., \citealt{caldwell.2003AJ....125.2891C}). On the other hand they are challenging to study because they are intrinsically faint and their metallicities are generally relatively low, so the lines are more difficult to measure.

dEs's appearance is mostly featureless, except for the bright ones, which have components such as disks, spiral arms, bars, lenses, irregular features (e.g. \citealt{jerjenetal.2000A&A...358..845J,barazzaetal.2002A&A...391..823B,gehaetal.2003AJ....126.1794G}; \citealt*{grahamandguzman.2003AJ....125.2936G}; \citealp{derijckeetal.2003A&A...400..119D,liskeretal.2006AJ....132..497L,ferrareseetal.2006ApJS..164..334F,janzetal.2012ApJ...745L..24J,janzetal.2014ApJ...786..105J,Suetal.2021A&A...647A.100S,Micheaetal.2022arXiv220506281M}). Contrary to giant E/S0 galaxies, their surface brightness profiles follow nearly exponential laws\footnote{https://unishare.nl/index.php/s/pxtKX534Wgo3Fcew} \citep{caonetal.1993MNRAS.265.1013C,ferrareseetal.2006ApJS..164..334F,liskeretal.2007ApJ...660.1186L}.  \citet{kormendy.1985ApJ...295...73K} suggested that they developed their spheroidal, non-star forming  and most likely highly flattened appearance
(\citealt{liskeretal.2006AJ....132..497L,liskeretal.2007ApJ...660.1186L}) during a transformation from a
late-type galaxy when falling into the cluster. In such a way this environmentally-induced transformation created the morphology-density relation (e.g. \citealt*{boselliandgavazzi_2014A&ARv..22...74B}), which is stronger for dwarfs than for giant galaxies. 

Over many years, unresolved stellar populations studies have been done comparing the observed spectrum with a combination of stellar populations models with a range of metallicity and age, and a limited range of abundance ratios. This allows obtaining the Star Formation History (SFH) (i.e, the distribution of stellar ages), the distribution of metallicities, and the abundances of specific key elements. In practise this is done by fitting the full spectrum to models or by measuring line indices, and fitting them to modelled values. Both methods have their own advantages and disadvantages. When using full spectrum fitting approaches one can measure SFHs (e.g. \citealt{cidfernandesetal.2005MNRAS.358..363C,kolevaetal.2009MNRAS.396.2133K}), but measuring abundance element ratios is not free from problems as there are uncertainties related to stellar atmospheric calculations. Going through line indices has the benefit that it allows obtaining several independent measurements of the abundance of each element and in this way have more control over the measurements. However, as this latter method is generally applied to low resolution spectral indices, only a few very strong lines and abundances of only a number of elements can be measured, and therefore loses the advantage of having many independent measurements of the same quantity. In here, we wanted to make use of our newly defined set of our high resolution indices.

 Using low- and medium resolution spectra, a widely used
method is to measure spectrophotometric indices, like Lick
\citep{worthey.1994ApJS...95..107W,Wortheyetal.2014A&A...561A..36W} or  Rose (1984) indices
and compare them to model predictions. Analysing the spectra of galaxies is tougher than it seems because of the age-metallicity degeneracy. Index–index 
diagram can be used to break the age-metallicity degeneracy in old
stellar populations. An optimised version of this age-sensitive
indicator is proposed in \citet{vazdekis.1999ApJ...513..224V,vazdekisetal.2001Ap&SS.276..839V}. \citet{jonesetal.1995ApJ...446L..31J} determined that the ${H}\gamma_{HR}$-Fe4668 diagram can be used to break the age-metallicity degeneracy in old stellar populations. \citet{wortheyandottaviani.1997ApJS..111..377W} discuss ${H}\delta$, a line that has lower sensitivity to metallicity, and ${H}\beta_{o}$ is the optimized ${H}\beta$ index, defined by \citet{cervantesandvazdekis.2009MNRAS.392..691C}, that is almost not metallicity dependent and therefore a better age indicator. Later on, researchers have continued using this system, since the velocity broadening in massive galaxies is so large that using a higher resolution system would not represent a significant advantage for those objects. For smaller stellar systems, the signal-to-noise ratios (S/N) in the data were often so low that a lower resolution system of indices was often preferred. However, for the smaller systems of this paper, much more information can be extracted if the higher resolution of the data is used. Here,
the broadening by stellar motion is so low, that many more lines are measurable than in giant galaxies. With a system of high resolution indices we get a better handle on the abundance distribution in these systems.

Integrated galaxy spectra contain the imprinted chemical evolution of these objects. For instance, massive elliptical galaxies are enhanced in [Mg/Fe] but lower-mass galaxies are not. More local observations indicate that the relative scarcity of low metallicity stars in the solar vicinity does not match predictions from simple galactic chemical evolution models ('G-dwarfs problem') in The Galaxy (e.g. \citealp{SearleandSargent.1972ApJ...173...25S}), besides the G-dwarf problem also appears in external galaxies like M31 \citep{wortheyetal.1996AJ....112..948W}. We know much less about the stellar populations of dwarf elliptical galaxies than about their more massive counterparts, which have been studied on an individual galaxy basis and by using surveys such as SDSS, Manga or SAMI.

The stellar populations of dE's span a wide range of subsolar metallicities, from [M/H] $\sim$ -0.1 to -1.5, and mean ages from 1 to $14$\,Gyr \citep{caldwell.2003AJ....125.2891C,michielsenetal.2008MNRAS.385.1374M,paudeletal.2010MNRAS.405..800P,kolevaetal.2011MNRAS.417.1643K,tolobaetal.2014ApJS..215...17T,rysetal.2015MNRAS.452.1888R,sybliskaetal.2017MNRAS,sybilskaetal.2018MNRAS.476.4501S,Sen.2018MNRAS.475.3453S}. Also, they do not consist of simple, old and metal rich stellar populations, but span a range in ages and are relatively metal poor systems \citep{michielsenetal.2008MNRAS.385.1374M, kolevaetal.2009MNRAS.396.2133K,rysetal.2015MNRAS.452.1888R}.

Stellar population studies show that unresolved dEs have on average a lower metal content than giant ellipticals, as expected from  the  metallicity-luminosity  relation \citep{michielsenetal.2008MNRAS.385.1374M,skillmanetal.1989ApJ...347..875S,sybliskaetal.2017MNRAS}. Their effective ages are somewhat younger on average, giving evidence of downsizing \citep{vazdekisetal.2004ApJ...601L..33V,nelanetal.2005ApJ...632..137N,thomasetal.2005ApJ...621..673T, sybliskaetal.2017MNRAS,caldwell.2003AJ....125.2891C}. However, recent studies show  that  the  stellar  populations  of  dEs  show  indications of both young and old ages and a range in gradients (e.g. \citealp{kolevaetal.2009MNRAS.396.2133K,kolevaetal.2011MNRAS.417.1643K,denbroketal.2011MNRAS.414.3052D,rysetal.2015MNRAS.452.1888R,hamrazetal.2019A&A...625A..94H}), in agreement with the few dEs in the Local Group. However studies about detailed abundance ratios in dEs are scarce. \citet{gorgasetal.1997ApJ...481L..19G,michielsenetal.2008MNRAS.385.1374M,sybliskaetal.2017MNRAS} and \citet{Sen.2018MNRAS.475.3453S}  show  that  [Mg/Fe]  is  around  solar, although not without scatter, lower  than  what  is  found  in  giant  ellipticals, with [Mg/Fe] possibly going up for decreasing metallicity. Very little is known about the abundance  ratios  for  other elements,  mainly because of the lack of high S/N spectra, but also because of the lack of methods and models to properly analyse them. In \citet{Sen.2018MNRAS.475.3453S} abundance ratios of [Ca/Fe] and [Na/Fe] are given for a sample of 39 Virgo dwarf ellipticals. [Ca/Fe] is found to be slightly larger than solar, while [Na/Fe] is considerably lower than solar. This is in sharp contrast with massive ETGs (see, e.g., \citealt{vazdekisetal.1997ApJS..111..203V} for Ca and \citealt{labarberaetal.2017MNRAS.464.3597L} for Na).

Most spectroscopy of dEs in the literature have been taken using long-slit instruments, and is therefore sensitive to aperture effects, the slit being almost always very narrow. With new Integral Field Spectroscopy this problem disappears, since a large part of a galaxy is covered in the integral-field unit (IFU). It is therefore much easier to produce high integrated S/N spectra than with long-slit spectroscopy, particularly for achieving high spectral resolution, which requires narrowing the slit.
Current IFU surveys (e.g. ATLAS 3D , SAMI, MANGA) do not include dEs due to low surface brightness and small size. Notably are the dwarfs observed with MANGA, and with stellar mass between 10$^9$ and 5.10$^9$ M$_\odot$ (\citealt{pennyetal.2016MNRAS.462.3955P, pennyetal.2018MNRAS.476..979P}) and the IFU observations using SAURON of $\sim$ 10 dEs by (\citealt{rysetal.2013MNRAS.428.2980R}). A high resolution study of the internal kinematics of dwarf galaxies in the Fornax Cluster, using observations of the SAMI instrument at a resolution of 5000, was recently published by \citet{scottetal.2020MNRAS.497.1571S}. Also, internal kinematics have been published of an infalling group of galaxies into the Virgo Cluster, by \citet{bidaranetal.2020MNRAS.tmp.2225B}, using the MUSE instrument. These studies have revealed that dEs are generally pressure supported stellar systems \citep{rysetal.2014MNRAS.439..284R,tolobaetal.2015ApJ...799..172T,scottetal.2020MNRAS.497.1571S,bidaranetal.2020MNRAS.tmp.2225B}, rotating even slower than their giant counterparts. 

This paper is designed as a two-part to understand the stellar populations of dwarf elliptical galaxies (dEs). In a first part we defined a system of high resolution absorption line strength, consisting of 23 indices measuring transitions of 12 elements. We investigated their behavior as a function of age, metallicity and [$\alpha$/Fe] abundance ratio. Here we aim at studying the processes that formed dEs by analysing their absorption line spectrum in detail. We will not use either classic Lick index or full spectral fitting approaches, but use the fact that dwarf ellipticals show only limited internal broadening, so that a newly defined system of high resolution absorption line indices can be used. In this way we can measure the indices of the system we defined in this paper. At the end of the paper, we will compare our results with other galaxies. These are mainly those in the Local Group, where individual stars have been observed, and those of giant ellipticals from integrated spectra. 

This paper is organised as follows. In Section 2 and 3, we describe the galaxy sample and the spectroscopic data we use in this paper, the instrumental setup used for the spectroscopic observations, and the main steps in the data reduction. In Section 4, we derive the mean luminosity weighted ages and metallicities of our galaxies based on age-sensitive and metallicity-sensitive indices. In Section 5, we introduce the new set of high-resolution line indices that we define and the method used for the definition and we characterise the dependence on age and metallicity of the indices and also investigate the dependence of the line indices on alpha-elemental abundance ratios. In Section 6, we present the measurement of the new set of high-resolution line indices for the dEs. In Section 7, we compare the galaxies with Pegase-HR SSP models, to find out how the abundance
ratio proxies compare with the ones in the solar neighborhood. In Section 7, we present the results, ordered by groups of elements and discuss them. 

\section{SAMPLE and OBSERVATIONS}
\begin{figure*}
	\includegraphics[width=0.99\textwidth]{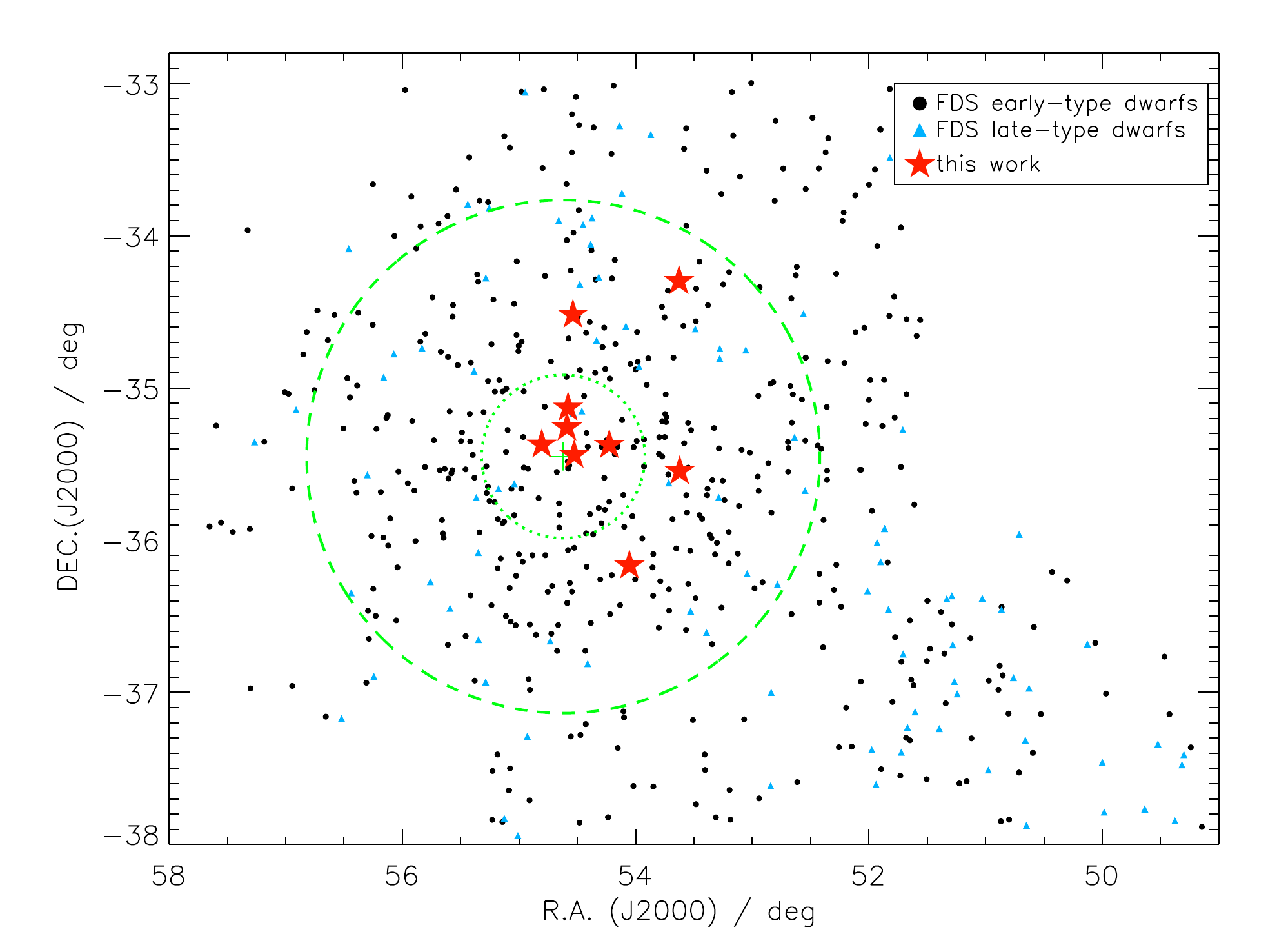}
   \caption{Map of the Fornax cluster. The black and blue symbols correspond to early-type dwarfs and late-type dwarfs from Venhola et al. (2018), respectively. The red stars represent our sample. The green dotted circle and lashed circle shows the core (Ferguson 1989) and  the virial radius of 2.2$\deg$ ($\sim$ 0.7 Mpc, Drinkwater et al. 2001) respectively. The green cross shows the central galaxy NGC 1399.}
    \label{ra_dec_fcc}
\end{figure*}

Our sample of dE galaxies is a subsample of the dEs observed by \citet{scottetal.2020MNRAS.497.1571S} using the SAMI IFU. This sample was selected from the FDS dwarf sample \citep{venholaetal.2018A&A...620A.165V}. For this paper we selected 8 dwarfs with high S/N spectra, stellar masses between 10$^8$ $M_{\sun}$ and 10$^{9.2}$ $M_{\sun}$, and with integrated velocity dispersions between 10 and 40 kms$^{-1}$. Venhola et al. classify them as dE and their properties (see Table 1) are representative for bright dwarf galaxies. Given that their effective radii range from 5 to 15 arcsec the spectra cover a region up to 0.5-1.5 ${R}_\textit{e}$ of the galaxy, appropriate for mapping stellar kinematics and extracting stellar populations representative of the whole galaxy. Three galaxies in our sample, FCC135, FCC182 and FCC203 are classified as disky from unsharp masking \citep{Micheaetal.2022arXiv220506281M}. Unsharp masking makes an effective way to examine if faint substructure features are embedded in the bright diffuse body of a galaxy. In \citet{Micheaetal.2022arXiv220506281M} disk structures show bars, spiral arms, rings and generally have the same colors as the galaxies, while clump substructures enclose irregular light over-densities such as star forming regions, dust lanes and off-center nuclei. None of our sample is classified as clumpy. In Fig.~\ref{ra_dec_fcc} we show the locations of our galaxies in Fornax cluster. They are all located within the Virial Radius, in the center of the cluster.

All observational data were obtained at the Sydney -- Australian Astronomical Observatory (AAO) Multi-Object Integral-Field spectrograph (SAMI; \citealp{croom.2012MNRAS.421..872C}) which is mounted at the prime focus of the 3.9m Anglo-Australian Telescope (AAT) at Siding Spring Observatory, New South Wales. 

SAMI is based on lightly fused fibre bundles called hexabundless \citep{bland-hawthornetal.2011OExpr..19.2649B,bryantetal.2011MNRAS.415.2173B,bryantetal.2014MNRAS.438..869B}. SAMI consists of 13 hexabundles, each having an on-sky diameter of 15\arcsec. Besides 13 hexabundles, SAMI also has 26 individual sky fibres which allows us sky subtraction for all IFU observations without the need to observe separate blank sky frames. SAMI can target 13 galaxies in a single observation, or more likely 12 galaxies and one standard calibration star, significantly decreasing the amount of time needed to build a large sample of galaxies with IFS data. This standard star is useful for several important steps in the data reduction (e.g. telluric correction, absolute flux calibration), in addition to allowing us to figure out the point spread function and transmission of each individual observation.

AAOmega is a double-armed spectrograph covering the blue and red optical spectral regions. AAOmega allows variable wavelength coverage and spectral resolution in each arm. Our configuration for SAMI uses the 1500V grating in the blue arm, giving R= $\lambda / \Delta\lambda$ $\sim 5000$ over the wavelength range 4660-5430 \AA .

For this work we are primarily interested in fitting the stellar absorption features covered by the blue arm of AAOmega, and therefore we analyse only this arm. This spectral range has been widely used for detailed stellar populations studies. In this work we are pushing forward its potential
by performing the analysis at higher resolution.

The observing strategy is taken from the SAMI Galaxy Survey \citep{Sharp.2015MNRAS.446.1551S}. The galaxies were observed on  4$^\mathrm{th}$ -- 8$^\mathrm{th}$ Nov 2015 and 26$^\mathrm{th}$ -- 30$^\mathrm{th}$ Oct 2016.
For each field we aimed to obtain 7 hours ($\sim 25,000$ seconds) of 
on-source integration time. Individual integrations were 
$\sim$ 1800\ s, with dithers of 0\farcs8 (half a fibre diameter) applied 
between exposures, following a 7-point hexagonal dither pattern, 
optimised for the SAMI hexabundles. The dither pattern ensures an even 
distribution of S/N over a hexabundle, accounting for the gaps between 
fibres. This 7-point dither pattern was repeated twice for each field, 
yielding $\simeq25,000$ second total exposure 
time per galaxy. Arc lamp calibrations and observations of primary 
spectophotometric standard stars from the European Southern Observatory 
Optical and UV Spectrophometric Standard Stars 
catalogue\footnote{Available at:\\ 
https://www.eso.org/sci/observing/tools/standards/spectra.html} were 
interspersed with the object exposures at regular intervals.

\begin{figure*}
	\includegraphics[width=0.99\textwidth]{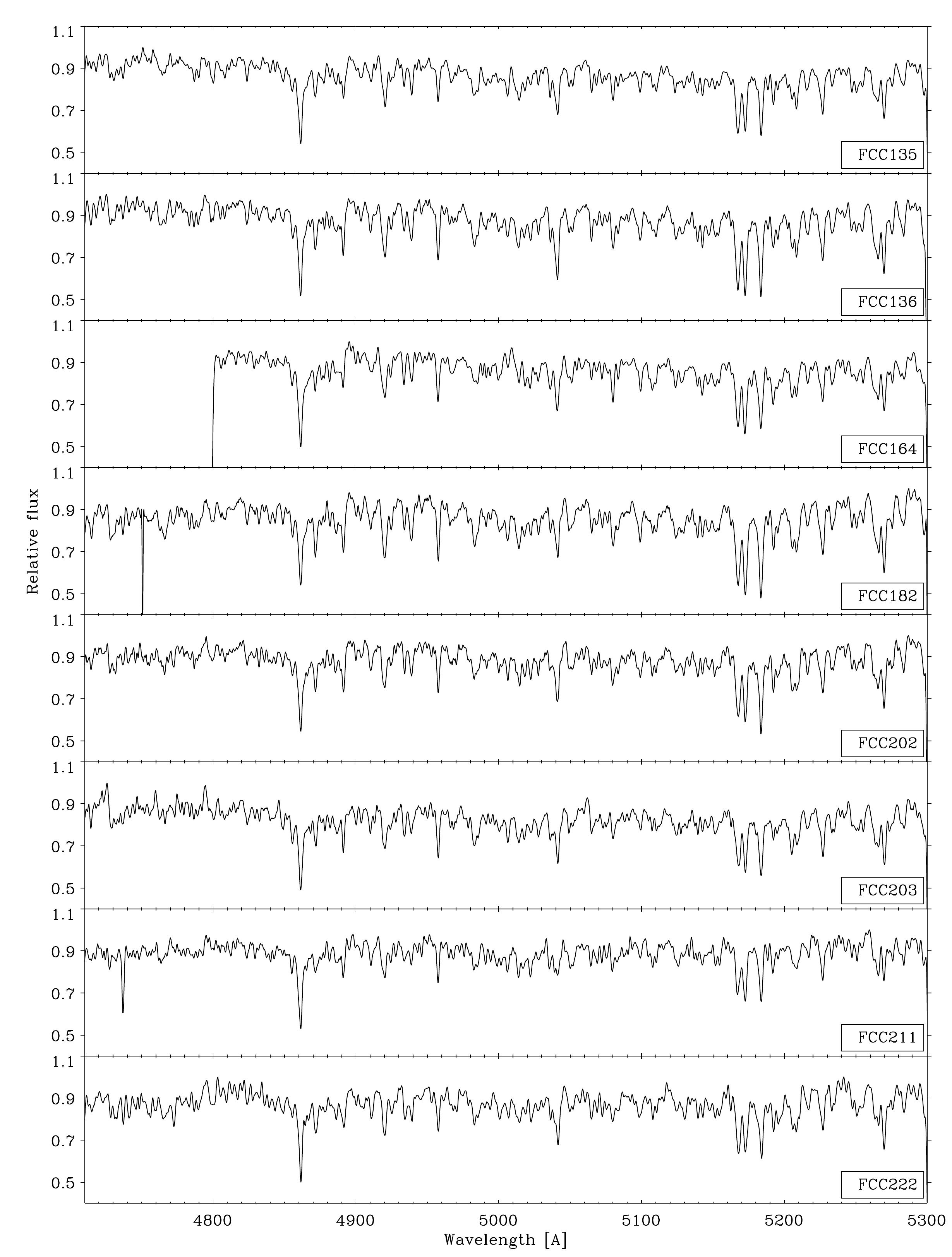}
   \caption{Normalized spectra of our 8 dEs taken with the SAMI.}
    \label{fig:spectra_sample}
\end{figure*}

The dEs galaxies were observed with SAMI using the 1500V grating in the blue arm. This grating provides a spectral resolution that is sufficient to resolve the typical velocity dispersion of dwarf galaxies of 20-30 km s$^{-1}$. In addition, it also covers the principal stellar absorption features (e.g. ${H}\beta$, Fe5015, Mgb) and the newly defined high-resolution line indices of Section 5, which were designed to measure relevant stellar population properties. Figure \ref{fig:spectra_sample} shows the normalized spectra of our sample dEs taken with the SAMI, identified by their names. 

\begin{table*}
 \caption{Properties of the dEs in Fornax Cluster .
  Column 1: galaxy name. Columns 2: FDSDC name by \citet{venholaetal.2018A&A...620A.165V}. Columns 3 and 4: right ascension and declination in J 2000. Columns 5 and 6: r-band and g-band magnitude (in the AB system), Columns 7: half-light radius by \citet{venholaetal.2018A&A...620A.165V}. Column 8: velocity dispersion from Eftekhari et al. (in preparation). Column 9 and 10: ellipticity and morphological classes from \citet{venholaetal.2018A&A...620A.165V}; e = smooth early-type, e* = smooth early-type, object has a nucleus, e(s) = early-types with structure, e(s)* = early-types with structure, object has a nucleus  Column 11: stellar mass. Column 12: date observed. }
 \label{tab:properties}
 \begin{tabular}{cccccccccccc} 
	    \hline
		Galaxy & FDS & RA  & Dec & M$_{r}$ &  M$_{g}$ & R$_{e}$ & $\sigma_\text{e}$ & $\epsilon$ & Morph. & log ($M\star /M_{\sun}$)& Date observed\\
		~   &   &   (deg)        & (deg)   & (mag)        & (mag)   & (arcsec)  & (kms$^{-1}$) &  &Class & & \\
		\hline
FCC135	& F15D384 &	53.628	&	-34.297	&	-16.8	&   -16.2	&   14.7	&	21.2	$\pm	2.8	$	&  0.53 & e(s)   &	8.708	$\pm	0.003	$	& Oct	2016	\\
FCC136	& F16D159 &	53.623	&	-35.546	&	-17.8	&   -17.0	&   17.5	&	30.9	$\pm	1.6	$	& 0.13	&  e &9.082	$\pm	0.003	$	& Oct	2016	\\
FCC164	& F12D367 &	54.054	&	-36.166	&	-16.0	&   -15.4	&   10.0	&	11.1	$\pm	5.2	$	& 0.45	& e(s)  &8.335	$\pm	0.003	$	& Oct	2016	\\
FCC182	& F11D279 &	54.226	&	-35.375	&	-17.9	&   -17.1	&   9.7	    &	38.9	$\pm	0.5	$	& 0.04	& e(s)*  &9.168	$\pm	0.002	$	& Oct	2016	\\
FCC202	& F11D235 &	54.527	&	-35.440	&	-17.3	&   -16.6	&   13.3	&	31.5	$\pm	1.0	$	&  0.41 & e*	& 8.909	$\pm	0.003	$	& Nov	2015	\\
FCC203	& F10D189 &	54.538	&	-34.519	&	-16.9	&   -16.3	&   16.0	&	31.4	$\pm	2.3	$	&  0.45	& e(s)   & 8.757	$\pm	0.003	$	& Oct	2016	\\
FCC211	& F11D339 &	54.590	&	-35.260	&	-16.1	&   -15.5	&   6.6	&	20.1	$\pm	5.7	$	&  0.25	& e*  & 8.339	$\pm	0.003	$	& Nov	2015	\\
FCC222	& F11D283 &	54.806	&	-35.371	&	-17.0	&    -16.3	&   16.1	&	18.6	$\pm	3.8	$	& 0.11& e*	& 8.771	$\pm	0.003	$	&  Nov 2015	\\

		\hline
  \end{tabular}
 \end{table*}
 
 \begin{figure*}
	\includegraphics[width=0.99\textwidth]{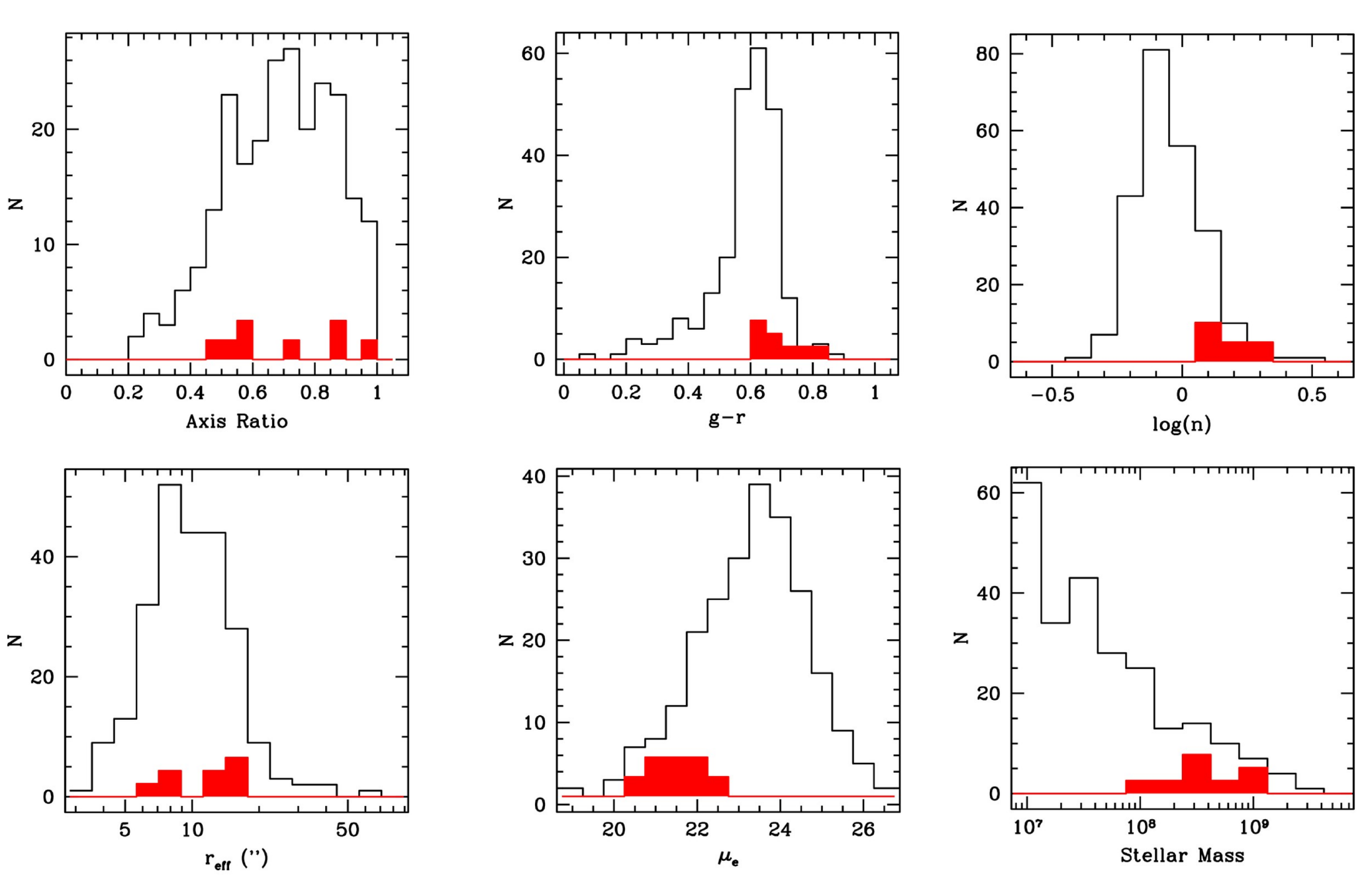}
   \caption{Properties of our sample, as compared to the sample of all dwarf galaxies more massive than 10$^7$ M$_\odot$ of the Fornax Cluster (Venhola et al. 2019, in black). The galaxies considered in this paper are plotted in filled red. }
    \label{fig:sample}
\end{figure*}

\section{DATA REDUCTION}
The reduction of the SAMI observations is described in \citet{Scott.2018MNRAS.481.2299S}, with further details provided in \citet{Sharp.2015MNRAS.446.1551S} 
and \citet{Green.2018MNRAS.475..716G}. Our SAMI data were reduced using the {\it sami} {\sc Python} package \citep{Allen.2015MNRAS.451.2780A}. Here we briefly summarise the process and give in detail the changes with respect to the previous works. 

The SAMI data reduction was performed in two main steps; the first takes the data from raw observed frames to Row-Stacked Spectra (RSS) frames, which is handled 
primarily to partially calibrated spectra from each
fibre of the instrument, including spectral extraction, the standard steps of bias subtraction, flat-fielding, wavelength calibration and sky subtraction by the two-degree field data reduction software package 
({\sc 2dfDR}\footnote{https://www.aao.gov.au/science/software/2dfdr}). The second step takes the data from RSS frames to flux-calibrated, three-dimensional data cubes, utilising purpose-built {\sc Python} software as part of the {\it sami} package. The entire process is overseen by the {\it sami} {\sc Python} {\it manager}.

Moreover, spectra corresponding to individual fibres are extracted  using `tramlines' fits to observations of the twilight sky. Subsequent to  the fibre extraction, telluric correction and relative and absolute flux  calibration steps are performed utilising the spectrophometric standard star and standard calibration star observations. Finally, the data for  each individual object are extracted from the RSS frames and combined into a three-dimensional data cube using a drizzle-based algorithm.

At the end, we add the spectra of the whole SAMI-field to get an integrated spectrum. More details  on  observations  and  data reduction are presented in \citet{scottetal.2020MNRAS.497.1571S} (S20).

\section{AGE and METALLICITY DERIVED USING STANDARD INDICATORS}

In this paper the process of determining abundance ratio proxies of various elements goes as follows. We first determine a mean luminosity weighted age and metallicity using standard age- and total metallicity-sensitive indicators (${H}\beta_\text{o}$ and [MgFe50]). After that we use a set of high resolution absorption line indices (see Section~\ref{sec:defining_indices}) to compare our galaxies with SSP model predicted indices for the derived age and metallicity. This gives an uncalibrated comparison, showing whether the index values for a certain galaxy are lower or higher than expected. These values are tabulated in Table~\ref{tab:abundance_proxy}. Given that models to calibrate such relative deviations are not available at the required resolution, we decided to only give these proxies, which themselves are very instructive and tell us about the way these galaxies were assembled. In this section we give details on how the ages and metallicities were calculated. In the next section the indices are defined.

To obtain the mean luminosity-weighted age and metallicity we assume that the galaxies are fitted with single-age and single-metallicity SSP models. This is done by measuring the age-sensitive (${H}\beta_\text{o}$) and metallicity-sensitive (Fe5015 and Mg $\textit{b}$) Lick spectral indices (\citealp{worthey.1994ApJS...95..107W}). Before measuring the indices we broadened our data out to $\sigma$ = 40 km\,s$^{-1}$ to match the PEGASE.HR (\citealp{leborgneetal.2004A&A...425..881L}) SSP models then
calculated the index values. We minimize the possible influence of abundance ratios on the derived SSP metallicity by using the abundance-ratio insensitive index combination [MgFe50] \citep{kuntschneretal.2010MNRAS.408...97K} and the ${H}\beta_\text{o}$ index \citep{cervantesandvazdekis.2009MNRAS.392..691C}, which is less dependent on metallicity than the standard Lick ${H}\beta$ index.

We use the grids of PEGASE.HR SSP models (with ages from 1 to 14 Gyr and [Fe/H] from -1.7 to 0.4) to match the data with stellar population models. Then, we obtain best-fit population parameters by effectively computing the “distance” from our measured indices to all the predicted values of those indices on the finer grid, and finding the age-metallicity combination with the minimum total distance. We estimate the ages and metallicities ([M/H]) using the software RMODEL
\citep{cardieletal.2003A&A...409..511C}. This software interpolates the age and metallicity inside an index-index grid. The errors in the age and [M/H] are calculated by running 1000 Monte Carlo simulations, varying the values of the spectral indices within a Gaussian function whose width is equal to their uncertainties.

\begin{figure}
\includegraphics[width=\columnwidth]{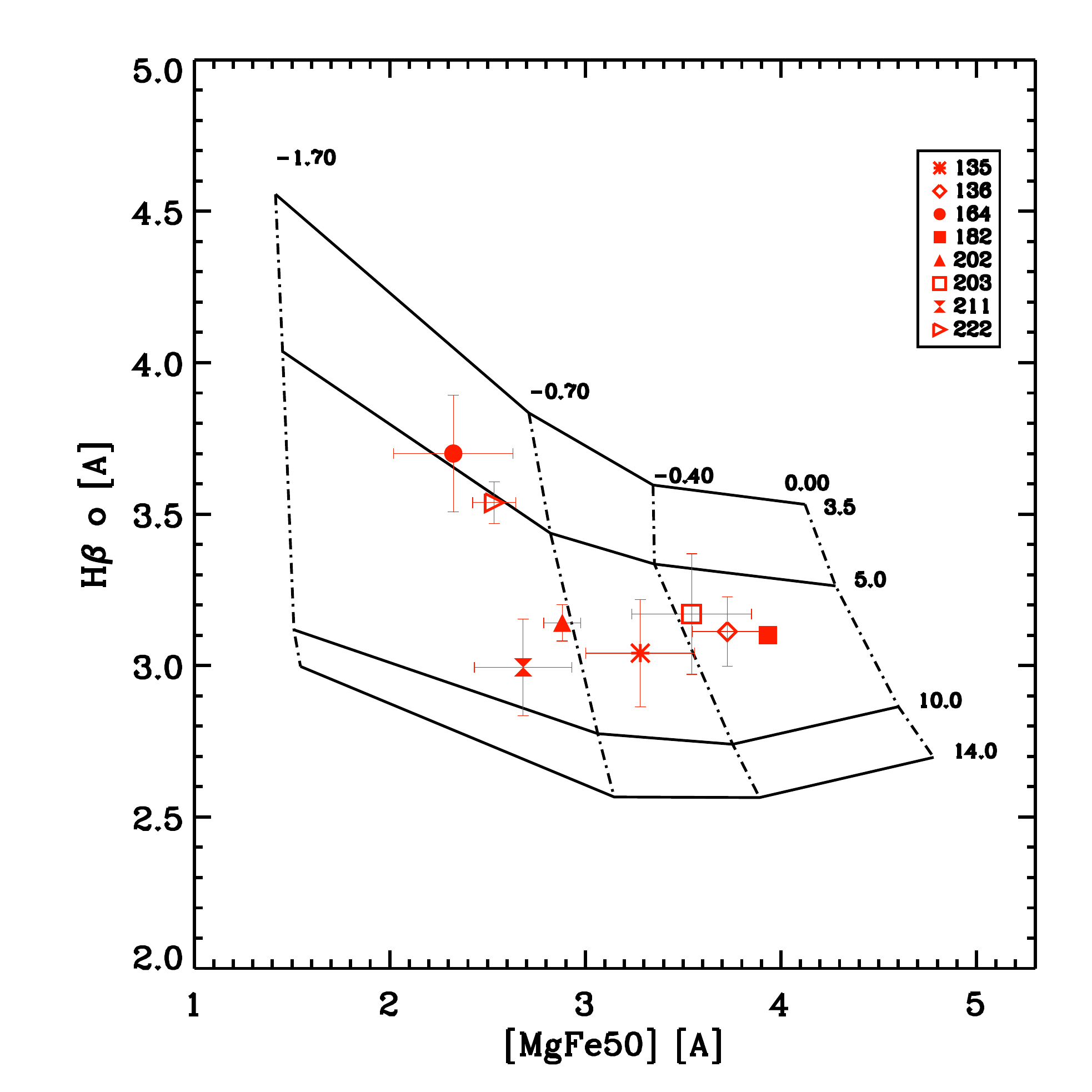}
\caption{${H}\beta_\text{o}$ vs. [MgFe50] line-strength, overplotted on the grid of PEGASE.HR (\citealp{leborgneetal.2004A&A...425..881L}) SSP models predictions with $\sigma$ = 40 km\,s$^{-1}$. Solid lines indicate constant age 3.5, 5.0, 10.0 and $14.0$\,Gyr, respectively, while dotted lines indicate constant [M/H] -1.70, -0.70, -0.40 and +0.00, respectively.}
\label{fig:index-index_hbo}
\end{figure}

\begin{figure}
\includegraphics[width=\columnwidth]{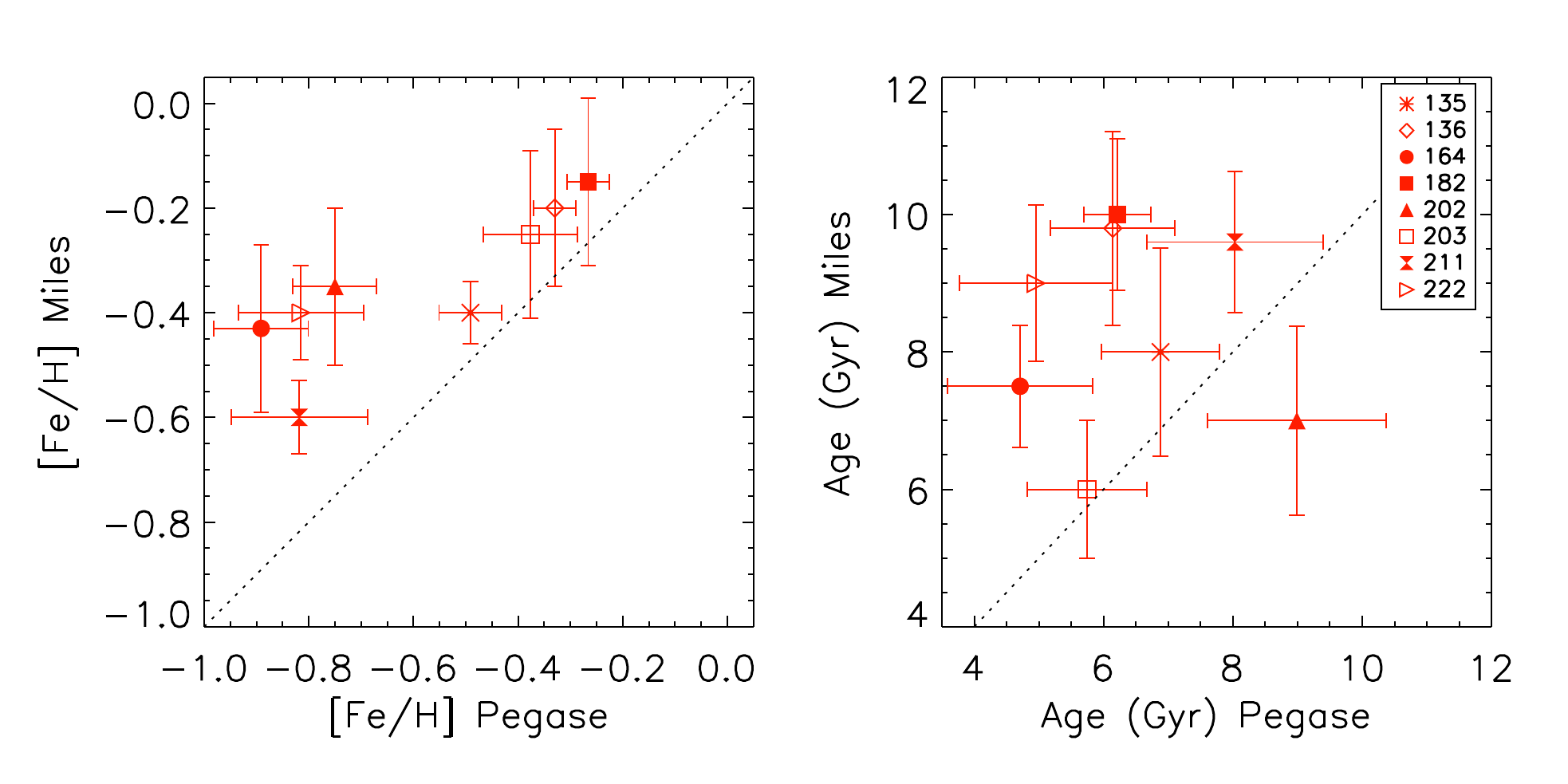}
\caption{Comparison between PEGASE.HR and MILES models with $\sigma$ = 60 km\,s$^{-1}$. }
\label{fig:comp_pg_mil}
\end{figure}

Fig.~\ref{fig:index-index_hbo} shows ${H}\beta_\text{o}$ as a function of the [MgFe50] index plots where we have restricted the age to the interval 3.5 - $14.0$\,Gyr, and the metallicity range from -1.70 to 0.00, which cover the range of solutions of our galaxies. Both luminosity weighted age and metallicity were computed, which are given in Table ~\ref{tab:indices_results}. We then compared with the MILES models \citep{vazdekisetal.2010MNRAS.404.1639V} after smoothing the data $\sigma$ = 60 km\,s$^{-1}$ , i.e. the nominal resolution of these models. The results are shown in Fig.~\ref{fig:comp_pg_mil}. We find a good agreement
between the results obtained with these two models, although a small
shift of about 0.1 dex is found for the derived metallicities and about
$1.0$\,Gyr for the mean ages.

We find that while the metallicity of these galaxies vary between [Fe/H]$\sim$ -0.2 to $\sim$ -0.9, while show a mean age $\sim$7.0\,Gyr. This explains why we can use an approach based on a single SSP. Note that if the mean luminosity-weighted ages were not as old we would require considering younger components on the top of the dominating old population to be able to properly fit the data.

Studies have shown that lower-luminosity early-type galaxies show a wider range in age than their more luminous counterparts (e.g. \citealp{wortheyandottaviani.1997ApJS..111..377W,kuntschneranddavis.1998MNRAS.295L..29K,caldwell.2003AJ....125.2891C})
Some studies find that the star formation histories of early-type galaxies depend on cluster properties and environment \citep{sanchez-blazquezetal.2003ApJ...590L..91S,trageretal.2000AJ....120..165T}. The Fornax cluster is more compact than the larger Virgo cluster, making it a good target for the study of environmental influences and obtaining a better picture of galaxy formation \citep{jordanetal.2007ApJS..171..101J}.

\citet{kuntschneretal.2000MNRAS.315..184K} concluded that all their ellipticals in Fornax cluster show a uniformly old age, forming a sequence in metallicity varying roughly from -0.25 to 0.30 in [Fe/H]. Furthermore, the S0 galaxies in his sample, which, at the same time, were usually of lower mass, generally had lower ages \citep{kuntschneranddavis.1998MNRAS.295L..29K}.

\section{A high resolution line index system}

Here we  present the definition of a new set of high-resolution Lick-type spectral indices, which make it possible to give a proxy on the abundance ratios in systems with low stellar velocity dispersion, such as dwarf galaxies, globular clusters, UDGs etc. Although some high-quality, high-resolution spectroscopic data for dEs is available, the problem is that current spectral indices are defined for low resolution spectra, which is not optimal for studying the fainter lines. To achieve this, we have used the ELODIE high resolution spectra \citep*{prugnielandsoubiran.2001A&A...369.1048P,prugnielandsoubrian.2004astro.ph..9214P} to define a set of new indices covering transitions involving as many elements as possible. Indices should depend as little as possible on spectral resolution, and, ideally,  be located in isolated patches of the spectrum with minimal crowding.

The ELODIE library is a stellar database of 1959 spectra for 1503 stars, observed with the echelle spectrograph ELODIE on the 193 cm telescope at the Observatoire de Haute Provence. The typical signal-to-noise (S/N) ratio of the spectra is 500 \AA$^{-1}$.  It has a large coverage of atmospheric parameters : $T_\text{eff}$ from 3000 K to 60000 K, log  $\textit{g}$ from -0.3 to 5.9 and [Fe/H] from -3.2 to +1.4. The library provides spectra at two different resolutions (R=42000 and R=10000), where we used R=10000 calibrated in physical flux ($\sigma$ = 12.7 km\,s$^{-1}$ at FWHM = 0.55 \AA).

We use a grid of single age and single metallicity (SSP) stellar population models computed with the evolutionary synthesis code: PEGASE.HR (\citealt{leborgneetal.2004A&A...425..881L}). These models are based on the empirical stellar library ELODIE.3 \citep*{prugnielandsoubiran.2001A&A...369.1048P,prugnielandsoubrian.2004astro.ph..9214P}. The SSP models have a range of -1.7 to 0.4 in metallicity [Fe/H] and 1 Myr to 20 Gyr in age. As the ELODIE stars follow the abundance pattern of the Milky-Way the resulting PEGASE.HR (\citealt{leborgneetal.2004A&A...425..881L}) models also follow this pattern.

\subsection{Index Definitions}
\label{sec:defining_indices}

Since the galaxy spectra that we analyse in this paper cover the range $\lambda\lambda$ 4700-5400 \AA, we identified major absorption features and continuum regions in this spectral range using the spectral atlas of Arcturus\footnote{\label{foot:spectroweb}\url{http://spectra.freeshell.org/spectroweb.html}}. Once selected, for each line we pre-defined three wavelength bands containing the feature and two pseudocontinua on both the blue and red sides of the central feature. The line indices are determined using their pseudo-equivalent width, which one gets by integrating the ratio between the pseudocontinuum and the flux over the central or index passband.

The lines are chosen from the Atlas of Arcturus to study as many elements as possible by defining indices around features which contain absorption lines from transitions of these elements, selecting lines that are as isolated as possible, together with continuum bands as devoid of lines as possible.

The initial definition provided us with a first look into the general behaviour of the index as a function of age, metallicity and velocity dispersion ($\sigma$).
After analysing this pre-definition, we improved it for each index. This is done by means of an automatic program that performs a multidimensional maximisation of index values. This programme changes the limits of the central feature and the two pseudocontinuum bands, around their initial values and for each new definition, it measures the value of the index for the whole set of models and finds the maximum value of the index. When doing this we made sure that the feature and continuum bands did not shift by more than 5 \AA\ from the original guess. After that we checked each lines by eye to make sure not including another absorption from a different element.

We measured the newly defined indices for the whole grid of models (with ages from 0.1 to 13 Gyr and metallicities from -1.7 to 0.4 [Fe/H]), at a velocity dispersion of  $\sigma$ $\sim$ 25 km\,s$^{-1}$, which is typical for these objects (Eftekhari et al., submitted to MNRAS). We also study the effect of changing the resolution by broadening the spectra with velocity dispersions out to $\sigma$ = 130 km\,s$^{-1}$. Figure~\ref{fig:final_subfig_v3_2022} shows the results for some indices, where we have plotted index versus [Fe/H] and age, respectively and their dependence on $\sigma$. All the other indices are shown in Figure~\ref{fig:final_subfig_v3_1_2022}-~\ref{fig:final_subfig_v3_3_2022} in the Apendix~B. In general all the index value decrease with increasing sigma but some being more stable than the others. Among the most stable indices in the covered velocity resolution range are Fe4920, Cr5265, Sc5083 and V4924.

We also measured the indices of Elodie stars with $\sigma$ = 25 km\,s$^{-1}$ (which corresponds to R=5000) to understand better the dependence on age and metalicity. All of line indices are shown in Figure~\ref{fig:3d_elodie_forpaper}-~\ref{fig:3d_elodie_4} in the Appendix. In Figure~\ref{fig:3d_elodie_forpaper} we show how the line strength indices change as a function of effective temperature, surface gravity, and metallicity. Remarkably there are a number of indices that are gravity sensitive like Cr4789, Cr5247 and V4924, all achieving larger index values for giant stars. We will investigate their potential to constrain the IMF in low velocity dispersion systems in a forthcoming paper.

We list the features that we defined in Table~\ref{tab:indexdefinitions}. The index name is composed of the element or elements that the index targets, along with the central wavelength of the index passband. Wavelengths are listed in units of angstroms. None of the features in Serven et al. is included, since they are meant for low resolution spectra. From the original Lick system we only include the Mg $\textit{b}$ index, since this is the only strong Mg-line in this part of the spectrum.

We tried to obtain at least one or more indices per element for as many elements as possible, with equivalent widths larger than 0.2 \AA\, in order to be able to measure the index well. For Fe and Ti there were so many lines that we only selected some of the brightest features. We also tried to select indices that were mostly dependent on age or metallicity, although most indices are dependent on both (Section~\ref{sec:dependence_alpha}).
 
 \subsection{Age and Metallicity Indicators}
\label{sec:age_metallicity_indicator}

In this section we characterise the dependence on age and metallicity of the indices. To do this, we calculate partial derivatives of the indices with regards to age and metallicity. For this we consider the indices ${i}$ for 25 representative SSP-models of different ages (${j=1,...,N_j=5}$) (t=0.1, 0.3, 1, 3, 13 Gyr) and metallicities (${k=1,...,N_k=5}$) ([Fe/H]= -1.7, -0.7, -0.4, 0.0, 0.4),  and estimate the slope of the iso-metal lines, $\beta$, and of the isochrones, $\alpha$, in those points. For each point equation~(\ref{eq:equation1}) and equation~(\ref{eq:equation2}) give the definitions for $\alpha_i$ and $\beta_i$, respectively, where I is the index name, Z is the metallicity [Fe/H], and t represents the age.

 \begin{equation}
    \alpha_i(t_j)=\Bigg|{{\frac{\delta{I_i(Z,t)} }{\delta{Z}}}}\Bigg|_{t_j}                              
   \qquad \qquad \qquad (j=1, . . . ,N_j)
	\label{eq:equation1}
\end{equation}

\begin{equation}
    \beta_i(Z_k)=\Bigg|{{\frac{\delta{I_i(Z,t)} }{\delta{\log{t}}}}}\Bigg|_{Z_k}                          
    \qquad \qquad \qquad  (k=1, ..., N_k)
	\label{eq:equation2}
\end{equation}

We now get the average, relative dependence of the index ${i}$ on metallicity by averaging the N${_j}$ (in this case 5) ages: 

\begin{equation}
    A_i =\sqrt{\frac{\sum_{j=1}^{N_j}{\left({\frac{\alpha_i(t_j)}{I_i}}\right)}^2}{N_j}}
	\label{eq:equationA}
\end{equation}
 and the dependence of the index ${i}$ on age is given by Equation~\ref{eq:equationB}.
\begin{equation}
    B_i =\sqrt{\frac{\sum_{k=1}^{N_k}{\left({\frac{\beta_i(Z_k)}{I_i}}\right)}^2}{N_k}}
	\label{eq:equationB}
\end{equation}

Figure~\ref{fig:final_a_and_b_elements_alpha} shows the values of A${_i}$ and B${_i}$ for each index ${i}$ grouped per element. Data points near the X-axis are metallicity-indicators, while the ones near the Y-axis are age indicators. We list the values of A${_i}$ and B${_i}$ for each index ${i}$ Table~\ref{tab:age_metallicity_dependence}. Most of our line indices are affected by metallicity and age. We indicate some indices that are more sensitive to age or metallicitiy as listed in Table~\ref{tab:age_metallicity_dependence}.

\subsection{Dependence on [$\alpha$/Fe]}
\label{sec:dependence_alpha}

The chemical abundance patterns in the spectra of stellar population are a direct tracer of histories of star formation and chemical enrichment of galaxies. So, in order to be able to measure not only metallicity but also abundance ratios of alpha elements we investigated the dependence of the line indices on alpha-elemental abundance ratios using the theoretical models of \citet{walcheretal.2009MNRAS.398L..44W} (hereafter W09). The SSP-spectra in the age range 2-13 Gyr, for abundance values [Fe/H] = -0.5 to 0.2 and [$\alpha$/Fe] = 0.0 and 0.4 at a constant resolution of FWHM = 1 \AA . This leaves us with powerful models to study high/medium resolution galaxy spectra. Note that these are fully theoretical models, i.e. based on the \citet{coelhoetal.2005A&A...443..735C} library of theoretical stars.

\begin{figure*}
	\includegraphics[width=0.98\textwidth]{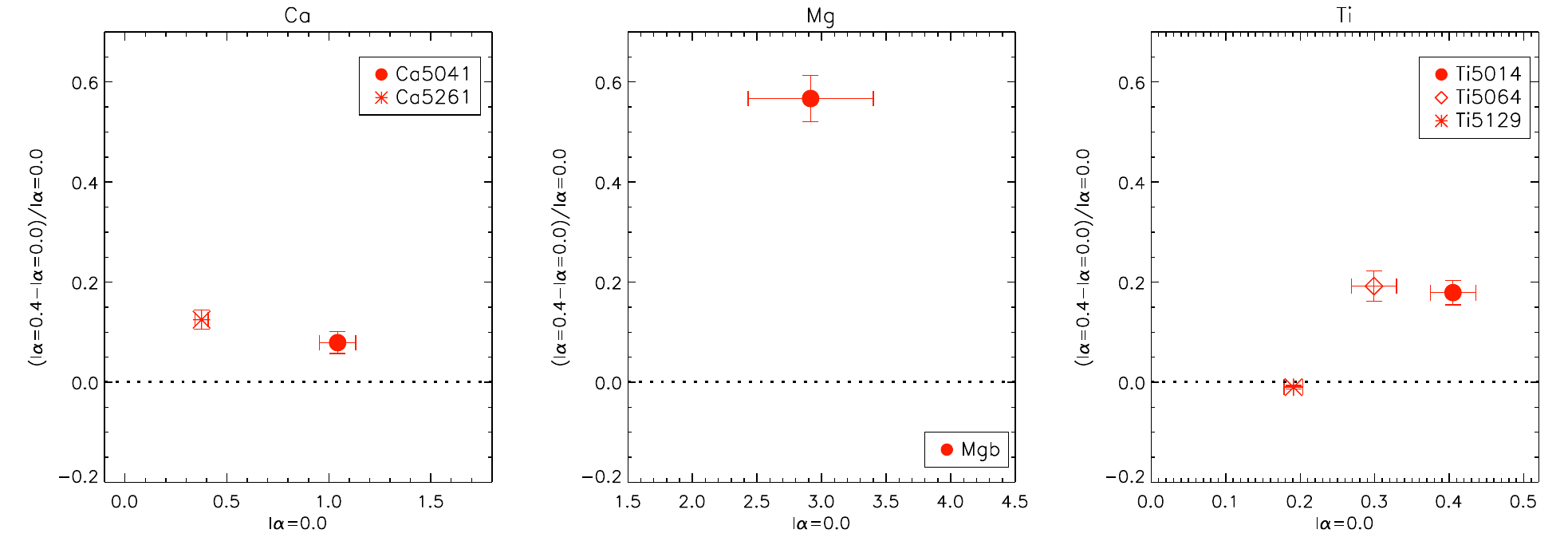}
    \caption{Differences between the alpha-enhanced predictions ([$\alpha$/Fe] = 0.4) and the solar-scaled ones ([$\alpha$/Fe] = 0.0) for $\alpha$-elements. On the x-axis is given the index at solar $\alpha$/Fe.}
    \label{fig:alpha_elemets_multi_witherror_alphaelements}
\end{figure*}

\begin{figure}
	\includegraphics[width=\columnwidth]{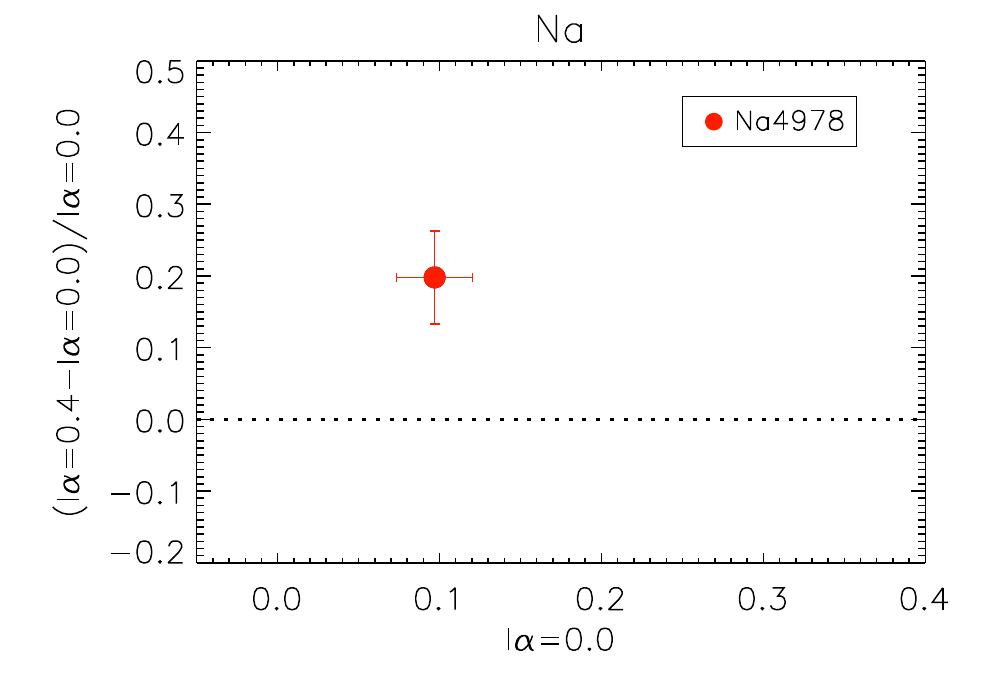}
    \caption{Differences between the alpha-enhanced predictions ([$\alpha$/Fe] = 0.4) and the solar-scaled ones ([$\alpha$/Fe] = 0.0) for the light odd-Z element: Na. On the x-axis is given the index at solar $\alpha$/Fe.}
    \label{fig:alpha_elemets_multi_witherror_Na}
\end{figure}

\begin{figure*}
	\includegraphics[width=0.98\textwidth]{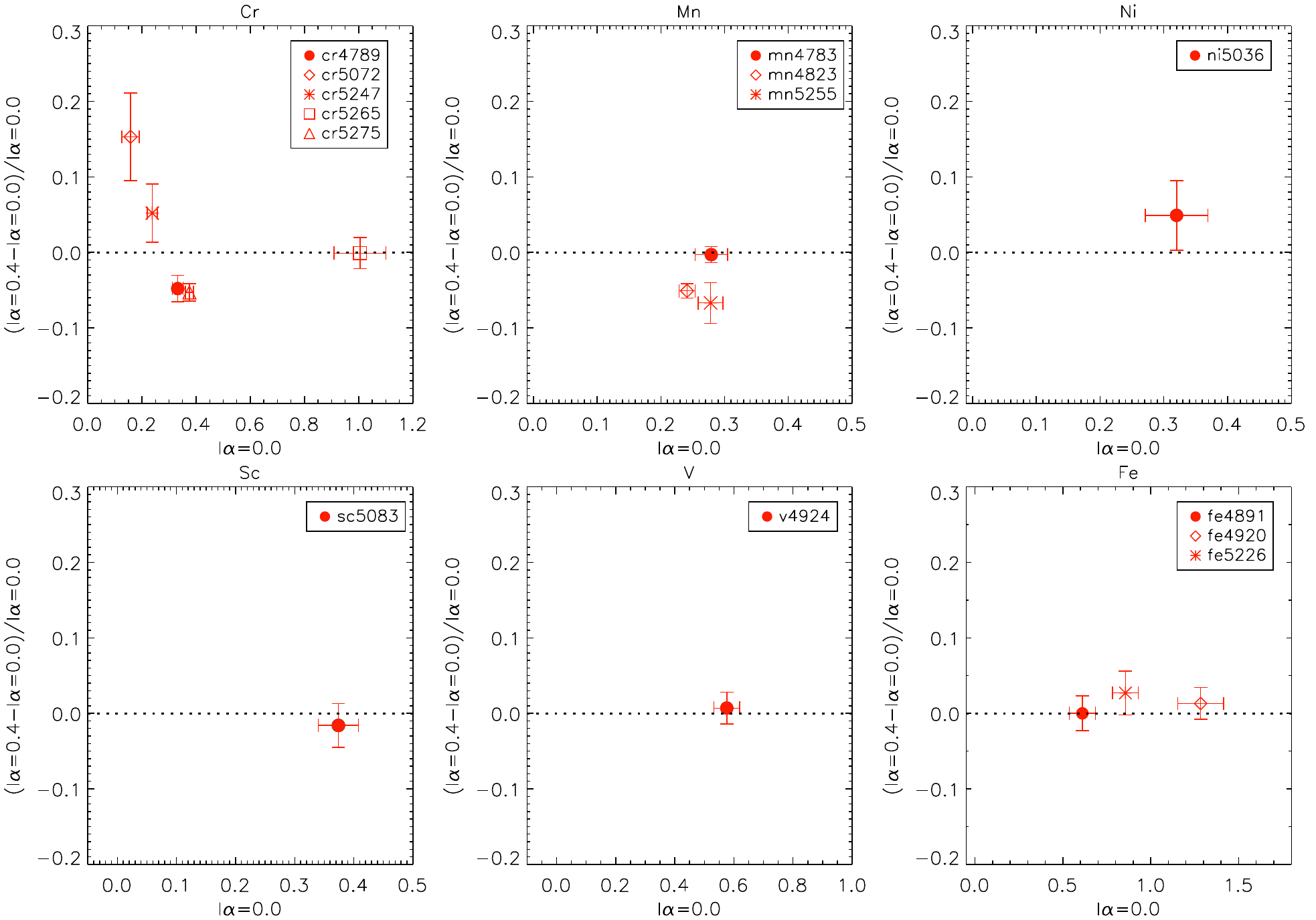}
    \caption{Differences between the alpha-enhanced predictions ([$\alpha$/Fe] = 0.4) and the solar-scaled ones ([$\alpha$/Fe] = 0.0) for Fe-peak elements. }
    \label{fig:alpha_elemets_multi_witherror_fepeak}
\end{figure*}

We compare the models of two different $\alpha$-element-to-iron abundance ratios ([$\alpha$/Fe] = 0.0 and 0.4) for the newly defined indices. 
When enhancing the abundances of the $\alpha$-elements, the abundances of the Fe-peak elements are decreased in order to keep the total metallicity fixed. 
Here we measure abundances of the $\alpha$-elements Mg, Ca and Ti (see Figure ~\ref{fig:alpha_elemets_multi_witherror_alphaelements}) and the Light odd-Z element: Na (see Figure~\ref{fig:alpha_elemets_multi_witherror_Na}) in addition to the Fe–peak elements Cr, Mn, Fe, Co, Ni, Sc and V (see Figure~\ref{fig:alpha_elemets_multi_witherror_fepeak}).

\begin{figure*}
	\includegraphics[width=0.98\textwidth]{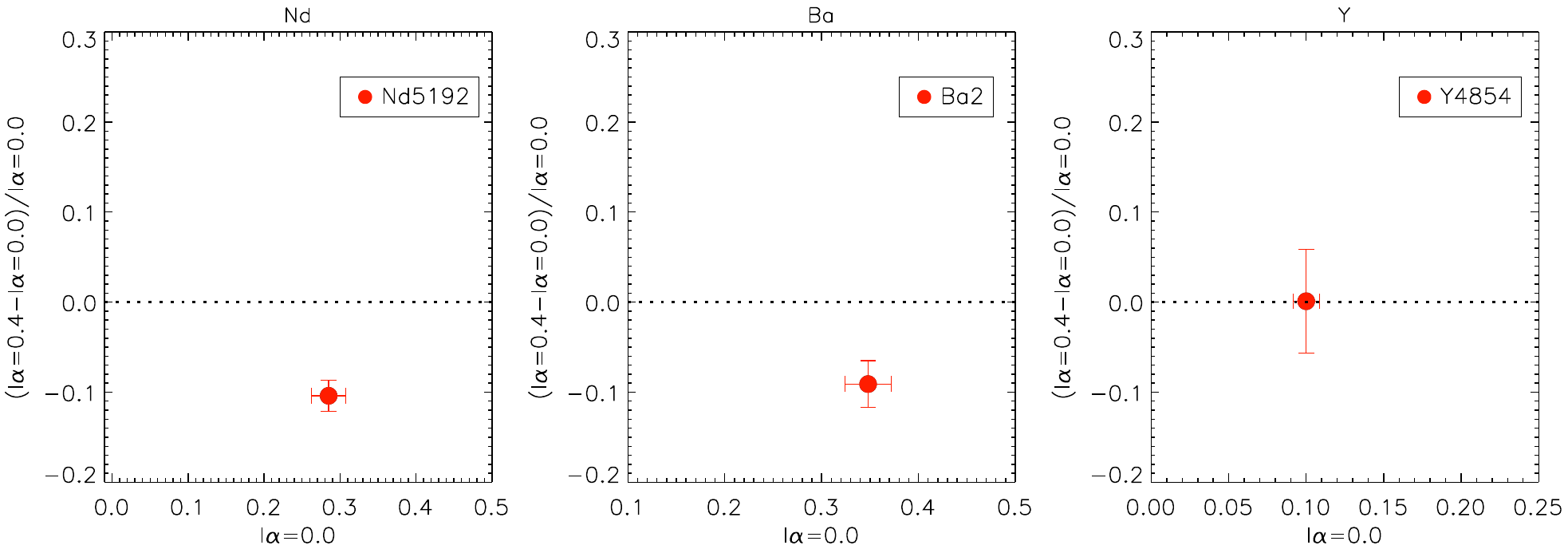}
    \caption{Differences between the alpha-enhanced predictions ([$\alpha$/Fe] = 0.4) and the solar-scaled ones ([$\alpha$/Fe] = 0.0) for heavy elements. }
    \label{fig:alpha_elemets_multi_witherror_heavyelements}
\end{figure*}

Indices characterised to be insensitive to $\alpha$/Fe can be defined as a good age or total metallicity indicators. Figures ~\ref{fig:alpha_elemets_multi_witherror_alphaelements} and ~\ref{fig:alpha_elemets_multi_witherror_fepeak} show that for the alpha enhanced models the $\alpha$-elements are seen to increase in strength, as expected, while the Fe-peak elements have similar strengths as for the models at solar abundance ratio, or are slightly weaker. The heavy elements shown in (Fig.~\ref{fig:alpha_elemets_multi_witherror_heavyelements} behave in the same way as the Fe-peak elements.


\section{Galaxy Measurements}
Unlike for massive galaxies, high resolution spectroscopy can provide accurate measurements of numerous absorption lines for many different chemical elements in dwarf galaxies, making it easier to derive their abundance pattern. Line indices are an effective way of analysing results as one can reduce the spectral data to simply a single number plus its error. In this work we study a new set of high-resolution spectral indices (Section~\ref{sec:defining_indices}), analogous to the Lick system in that the indices are defined by means of a blue and a red bandpass at either side of the feature bandpass. The given name corresponds to the element that the index targets, along with the central wavelength of the feature bandpass.

SAMI utilised the 1500V grating in the blue arm, giving R$\sim$5000 over the wavelength range 4660-5430 \AA. The indices were calculated using the package REDUCEME (\citealp{cardiel.1999PhDT........12C}).  For the interpretation of these high-resolution indices we focus on 23 indices which show good correlation with mass to study the abundance measurements.

For our galaxies the index measurements with errors in angstroms of equivalent width which we used in this work are given in Table \ref{tab:indices_results}.
\begin{table*}
 \caption{High-resolution spectral indices (as defined in this paper), age and metallicity, measured for 8 dEs in Fornax cluster }
 \label{tab:indices_results}
 \begin{tabular}{lcccccccc} 
	    \hline \hline
	 ~	 & FCC135  & FCC136  & FCC164  & FCC182 &FCC202 &FCC203&FCC211&FCC222 \\

		\hline \hline
log age  & 0.84	$\pm	0.03	$ &	0.79	$\pm	0.03	$ &	0.67	$\pm	0.04 $& 0.79	$\pm	0.02	$ &0.95	$\pm	0.07	$	& 0.76	$\pm	0.03	$	& 0.90	$\pm	0.05	$	& 0.70	$\pm	0.06	$ \\
(Gyr) & \\

[Fe/H] & -0.49	$\pm	0.06	$& -0.33	$\pm	0.04	$ & -0.89	$\pm	0.09	$& -0.27	$\pm	0.04	$& -0.75	$\pm	0.08	$& -0.38	$\pm	0.09	$& -0.82	$\pm	0.13 $& -0.81	$\pm	0.12	$ \\
\hline
Indices  & \\
\hline
Ba2	&	0.361	$\pm	0.056	$	&	0.409	$\pm	0.036	$	&	0.333	$\pm	0.060	$	&	0.401	$\pm	0.016	$	&	0.274	$\pm	0.019	$	&	0.219	$\pm	0.066	$	&	0.320	$\pm	0.051	$	&	0.355	$\pm	0.022	$	\\
Ca5041	&	0.991	$\pm	0.074	$	&	1.454	$\pm	0.045	$	&	0.786	$\pm	0.083	$	&	0.944	$\pm	0.021	$	&	0.945	$\pm	0.025	$	&	1.108	$\pm	0.081	$	&	0.742	$\pm	0.065	$	&	1.015	$\pm	0.028	$	\\
Ca5261	&	0.275	$\pm	0.042	$	&	0.387	$\pm	0.026	$	&	0.368	$\pm	0.043	$	&	0.359	$\pm	0.011	$	&	0.314	$\pm	0.013	$	&	0.327	$\pm	0.043	$	&	0.253	$\pm	0.035	$	&	0.215	$\pm	0.015	$	\\
Cr4789	&	0.229	$\pm	0.049	$	&	0.353	$\pm	0.030	$	&	. . .	&	0.378	$\pm	0.014	$	&	0.261	$\pm	0.016	$	&	0.338	$\pm	0.050	$	&	0.245	$\pm	0.043	$	&	0.314	$\pm	0.018	$	\\
Cr5072	&	0.134	$\pm	0.030	$	&	0.159	$\pm	0.019	$	&	0.115	$\pm	0.032	$	&	0.191	$\pm	0.008	$	&	0.158	$\pm	0.010	$	&	0.177	$\pm	0.033	$	&	0.130	$\pm	0.027	$	&	0.180	$\pm	0.012	$	\\
Cr5247	&	0.231	$\pm	0.041	$	&	0.245	$\pm	0.027	$	&	0.175	$\pm	0.044	$	&	0.262	$\pm	0.011	$	&	0.214	$\pm	0.013	$	&	0.125	$\pm	0.044	$	&	0.194	$\pm	0.035	$	&	0.194	$\pm	0.015	$	\\
Cr5265	&	0.605	$\pm	0.048	$	&	1.001	$\pm	0.028	$	&	0.692	$\pm	0.050	$	&	1.035	$\pm	0.012	$	&	0.813	$\pm	0.014	$	&	0.800	$\pm	0.047	$	&	0.620	$\pm	0.040	$	&	0.677	$\pm	0.017	$	\\
Cr5275	&	0.268	$\pm	0.048	$	&	0.406	$\pm	0.030	$	&	0.246	$\pm	0.051	$	&	0.404	$\pm	0.013	$	&	0.342	$\pm	0.015	$	&	0.331	$\pm	0.049	$	&	0.258	$\pm	0.041	$	&	0.292	$\pm	0.017	$	\\
Fe4891	&	0.504	$\pm	0.047	$	&	0.612	$\pm	0.030	$	&	0.472	$\pm	0.049	$	&	0.708	$\pm	0.013	$	&	0.617	$\pm	0.016	$	&	0.393	$\pm	0.056	$	&	0.330	$\pm	0.045	$	&	0.515	$\pm	0.019	$	\\
Fe4920	&	0.944	$\pm	0.083	$	&	1.257	$\pm	0.052	$	&	1.116	$\pm	0.086	$	&	1.389	$\pm	0.022	$	&	0.968	$\pm	0.028	$	&	0.778	$\pm	0.097	$	&	0.711	$\pm	0.077	$	&	0.928	$\pm	0.032	$	\\
Fe5226	&	0.812	$\pm	0.093	$	&	0.898	$\pm	0.059	$	&	0.767	$\pm	0.100	$	&	0.931	$\pm	0.025	$	&	0.771	$\pm	0.029	$	&	0.865	$\pm	0.099	$	&	0.520	$\pm	0.081	$	&	0.831	$\pm	0.033	$	\\
Mg $\textit{b}$ & 2.288$\pm	0.272$&	2.822$\pm	0.173$&	2.213$\pm	0.294$&	3.367$\pm	0.037$&	2.495$\pm	0.087$&	2.097$\pm	0.297$&	1.984$\pm	0.228$&	1.858$\pm	0.102$\\
Mn4783	&	0.152	$\pm	0.049	$	&	0.324	$\pm	0.030	$	&	. . .	&	0.269	$\pm	0.014	$	&	0.173	$\pm	0.017	$	&	0.221	$\pm	0.051	$	&	0.233	$\pm	0.042	$	&	0.140	$\pm	0.019	$	\\
Mn4823	&	0.183	$\pm	0.039	$	&	0.196	$\pm	0.025	$	&	0.004	$\pm	0.044	$	&	0.237	$\pm	0.011	$	&	0.183	$\pm	0.013	$	&	0.239	$\pm	0.042	$	&	0.146	$\pm	0.035	$	&	0.104	$\pm	0.015	$	\\
Mn5255	&	0.214	$\pm	0.047	$	&	0.309	$\pm	0.029	$	&	0.266	$\pm	0.048	$	&	0.292	$\pm	0.012	$	&	0.260	$\pm	0.015	$	&	0.288	$\pm	0.048	$	&	0.153	$\pm	0.039	$	&	0.300	$\pm	0.017	$	\\
Na4978	&	0.102	$\pm	0.043	$	&	0.120	$\pm	0.028	$	&	-0.002	$\pm	0.048	$	&	0.132	$\pm	0.012	$	&	0.013	$\pm	0.015	$	&	0.071	$\pm	0.050	$	&	0.075	$\pm	0.038	$	&	-0.058	$\pm	0.018	$	\\
Nd5192	&	0.273	$\pm	0.031	$	&	0.297	$\pm	0.020	$	&	0.340	$\pm	0.034	$	&	0.351	$\pm	0.008	$	&	0.300	$\pm	0.010	$	&	0.180	$\pm	0.034	$	&	0.229	$\pm	0.027	$	&	0.279	$\pm	0.012	$	\\
Ni5036	&	0.350	$\pm	0.048	$	&	0.429	$\pm	0.030	$	&	0.261	$\pm	0.054	$	&	0.456	$\pm	0.014	$	&	0.279	$\pm	0.017	$	&	0.400	$\pm	0.053	$	&	0.301	$\pm	0.042	$	&	0.265	$\pm	0.019	$	\\
Sc5083	&	0.184	$\pm	0.041	$	&	0.350	$\pm	0.025	$	&	0.177	$\pm	0.043	$	&	0.403	$\pm	0.011	$	&	0.328	$\pm	0.013	$	&	0.235	$\pm	0.044	$	&	0.259	$\pm	0.035	$	&	0.223	$\pm	0.016	$	\\
Ti5014	&	0.379	$\pm	0.043	$	&	0.460	$\pm	0.027	$	&	0.270	$\pm	0.044	$	&	0.441	$\pm	0.012	$	&	0.374	$\pm	0.014	$	&	0.388	$\pm	0.047	$	&	0.331	$\pm	0.041	$	&	0.318	$\pm	0.017	$	\\
Ti5064	&	0.264	$\pm	0.044	$	&	0.375	$\pm	0.027	$	&	0.090	$\pm	0.049	$	&	0.312	$\pm	0.012	$	&	0.250	$\pm	0.015	$	&	0.331	$\pm	0.048	$	&	0.229	$\pm	0.039	$	&	0.295	$\pm	0.017	$	\\
Ti5129	&	0.124	$\pm	0.036	$	&	0.181	$\pm	0.023	$	&	0.112	$\pm	0.038	$	&	0.203	$\pm	0.010	$	&	0.210	$\pm	0.012	$	&	0.128	$\pm	0.039	$	&	0.110	$\pm	0.030	$	&	0.169	$\pm	0.013	$	\\
V4924	&	0.481	$\pm	0.054	$	&	0.526	$\pm	0.034	$	&	0.421	$\pm	0.058	$	&	0.605	$\pm	0.015	$	&	0.384	$\pm	0.019	$	&	0.333	$\pm	0.063	$	&	0.306	$\pm	0.049	$	&	0.469	$\pm	0.021	$	\\
Y4854	&	0.128	$\pm	0.041	$	&	0.076	$\pm	0.027	$	&	0.127	$\pm	0.045	$	&	0.143	$\pm	0.012	$	&	0.101	$\pm	0.014	$	&	0.196	$\pm	0.045	$	&	0.137	$\pm	0.037	$	&	0.109	$\pm	0.016	$	\\
H\textit{{$\beta_o$}}	& 3.041$\pm	0.177$	 &3.112$\pm	0.115$	&3.700$\pm	0.193$	&3.101$\pm	0.026$	&3.141$\pm	0.060$	&3.170$\pm	0.199$	&2.994$\pm	0.159$	&3.538$\pm	0.069$\\

Fe5015 & 4.983$\pm	0.524$	&5.505$\pm	0.337$	&3.123$\pm	0.575$	&5.545$\pm	0.074$	&4.042$\pm	0.178$	&5.641$\pm	0.577$	&3.996$\pm	0.472$	&3.785$\pm	0.208$\\
		\hline
  \end{tabular}
 \end{table*}

\subsection{Comparison of data with stellar population models}

We use the single age and metallicity (SSP) stellar population models computed with the evolutionary synthesis code: PEGASE.HR (\citealp{leborgneetal.2004A&A...425..881L}). These models are based on the empirical stellar library ELODIE.3 (\citealp{prugnielandsoubiran.2001A&A...369.1048P,prugnielandsoubrian.2004astro.ph..9214P}) which has a range of -1.7 to 0.4 in metallicity [Fe/H] and 1 Myr to $20$\,Gyr in age.
 
We convolved all galaxies and all models to the same resolution: the PEGASE.HR models were convolved to an instrumental resolution of $\sigma$= 25 kms$^{-1}$(which corresponds to R=5000, and is the same as the SAMI-resolution), after which all models and all galaxies were convolved to an instrumental velocity dispersion of  $\sigma$= 40 kms$^{-1}$. This way all galaxies could be compared at once with each other.

\begin{figure*}
	\includegraphics[width=0.99\textwidth]{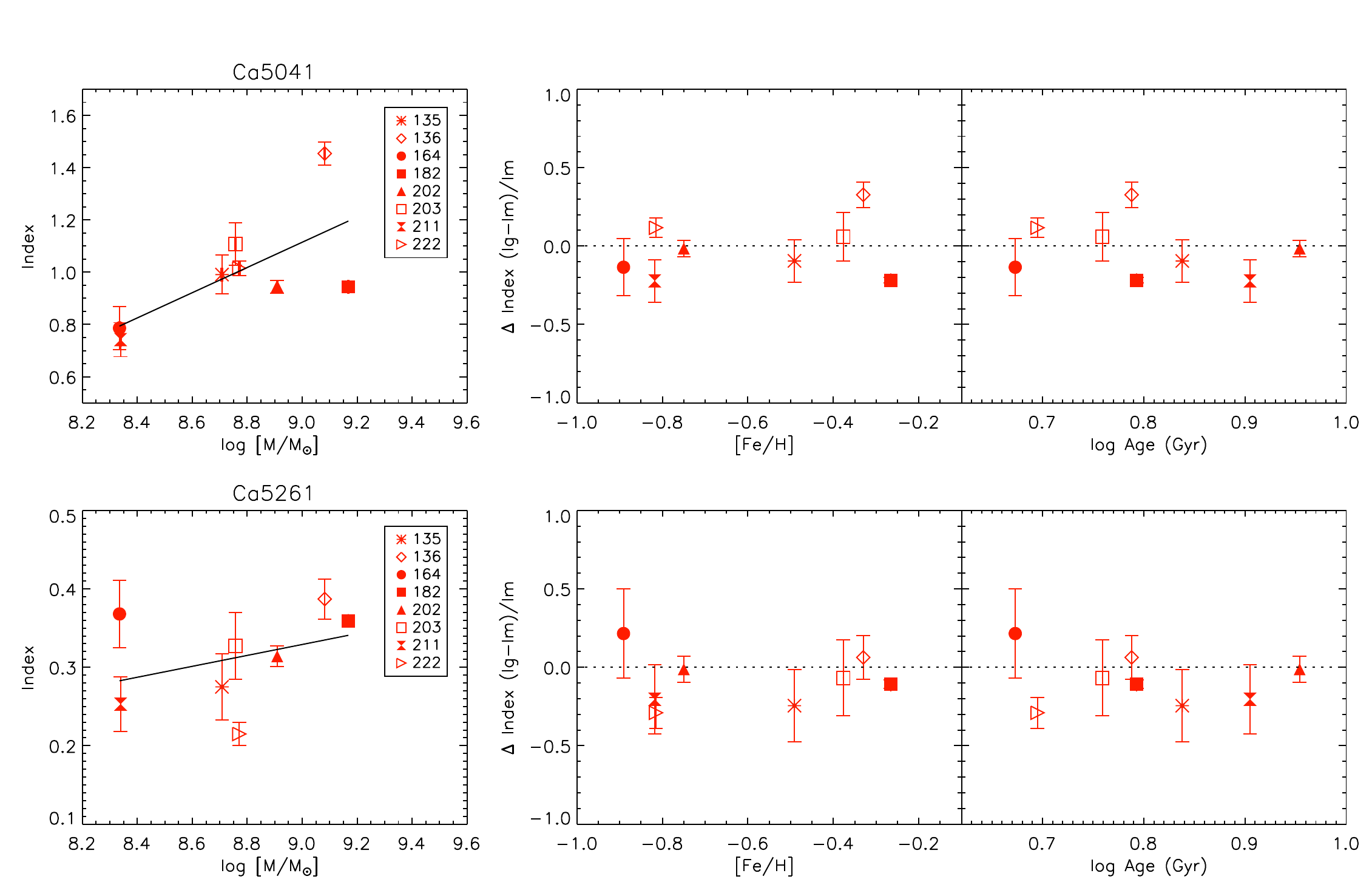}
   \caption{Comparison between index values of PEGASE.HR models and galaxies newly defined Ca lines are plotted as a function of galaxy mass (left panel), are shown in the center and right panels.}
    \label{fig:final_subfig_diff_norm_wg_ca_new}
\end{figure*}

\begin{figure*}
	\includegraphics[width=0.99\textwidth]{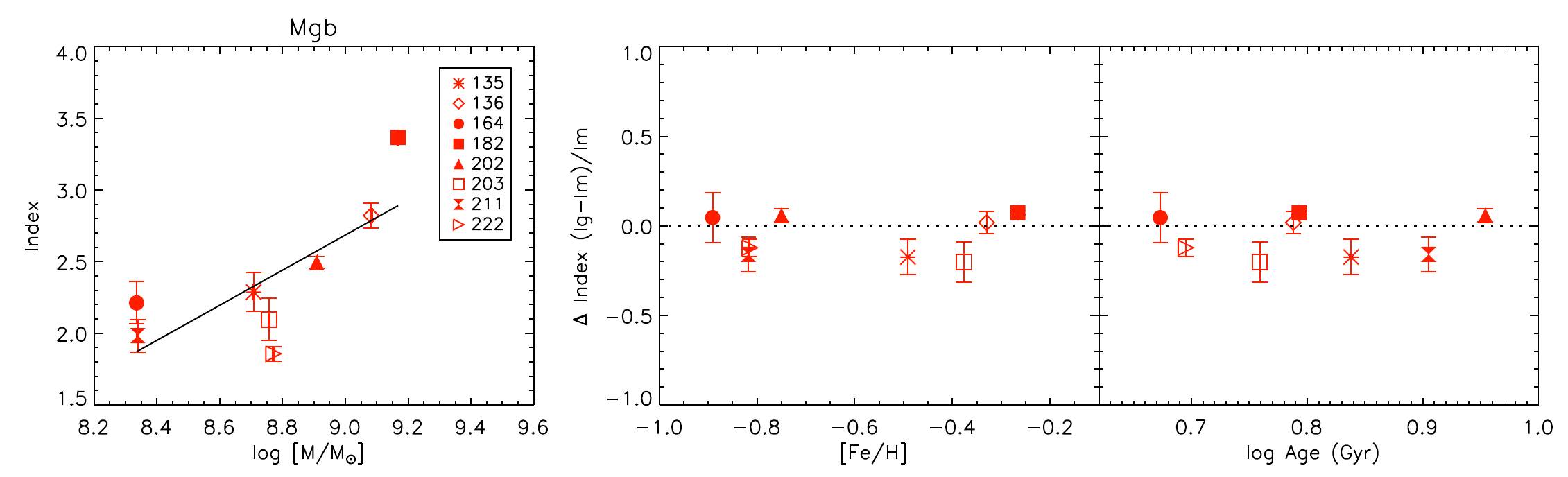}
   \caption{Comparison between index values of PEGASE.HR models and galaxies Mg line is plotted as a function of galaxy mass (left panel), is shown in the center and right panels.}
    \label{fig:final_subfig_diff_norm_wg_mg_new}
\end{figure*}

\begin{figure*}
	\includegraphics[width=0.99\textwidth]{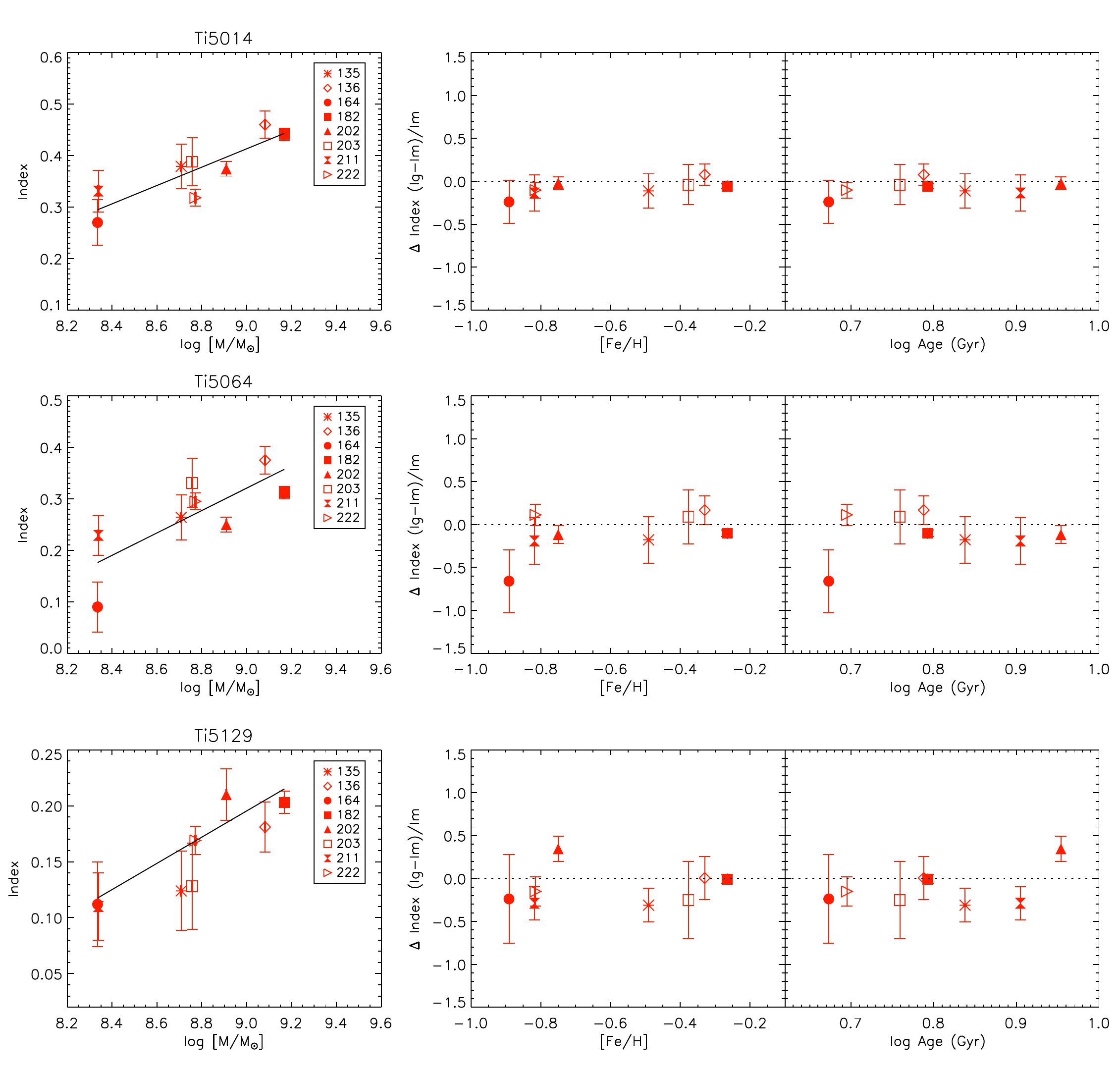}
   \caption{Comparison between index values of PEGASE.HR models and galaxies newly defined Ti lines are plotted as a function of galaxy mass (left panel), are shown in the center and right panels.}
   \label{fig:final_subfig_diff_norm_wg_ti_new}
\end{figure*}

\begin{figure*}
	\includegraphics[width=0.99\textwidth]{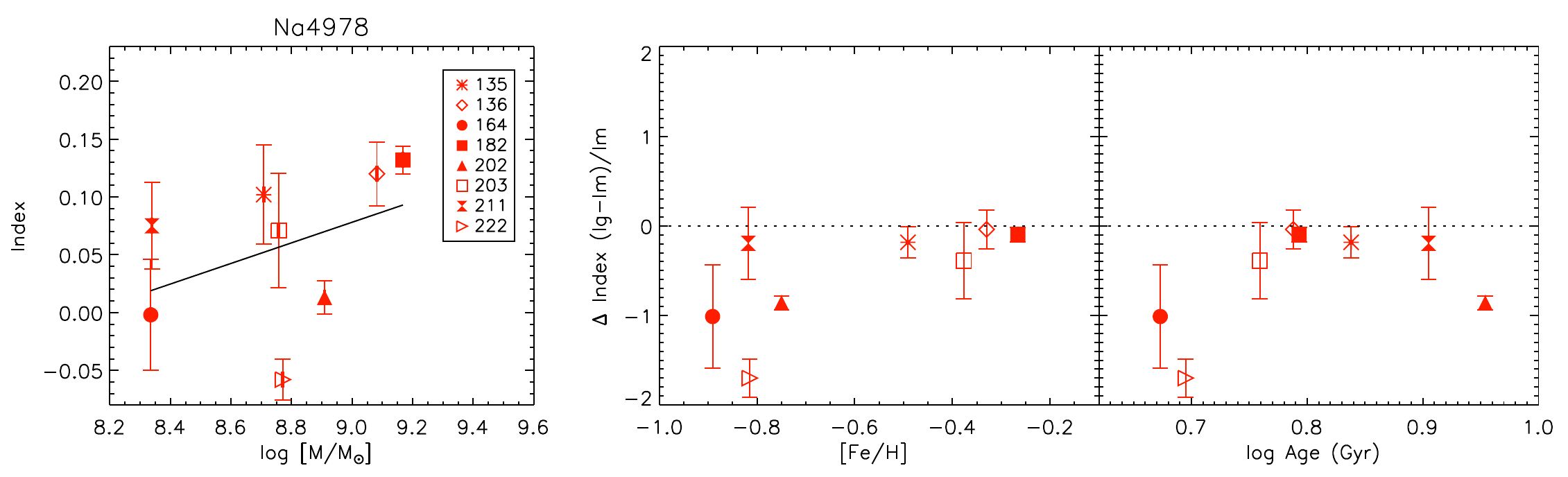}
   \caption{Comparison between index values of PEGASE.HR models and galaxies newly defined Na line is plotted as a function of galaxy mass (left panel), is shown in the center and right panels.}
    \label{fig:final_subfig_diff_norm_wg_na_new}
\end{figure*}

\begin{figure*}
	\includegraphics[width=0.99\textwidth]{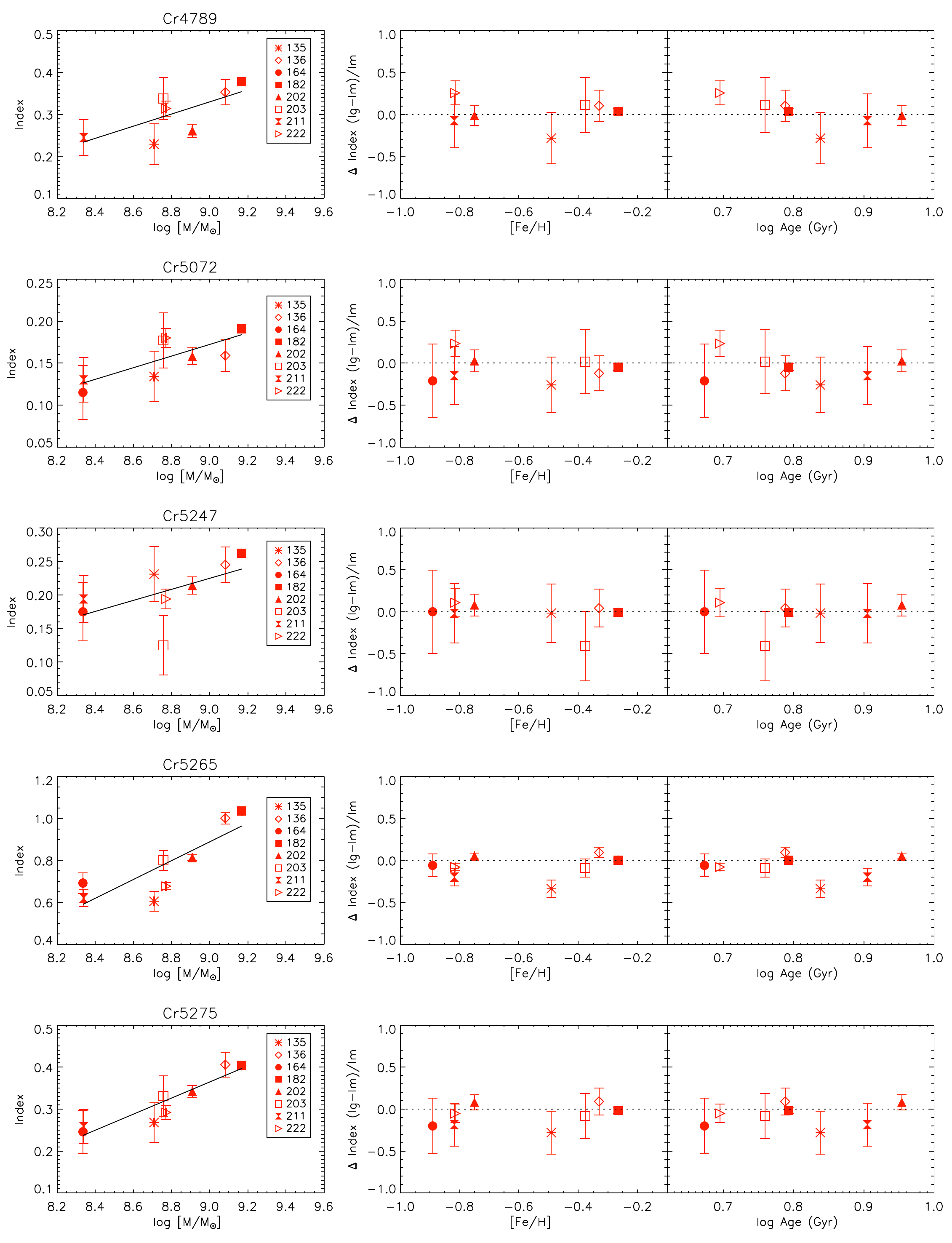}
   \caption{Comparison between index values of PEGASE.HR models and galaxies newly defined Cr lines are plotted as a function of galaxy mass (left panel), are shown in the center and right panels.}
    \label{fig:final_subfig_diff_norm_wg_cr1}
\end{figure*}

\begin{figure*}
	\includegraphics[width=0.99\textwidth]{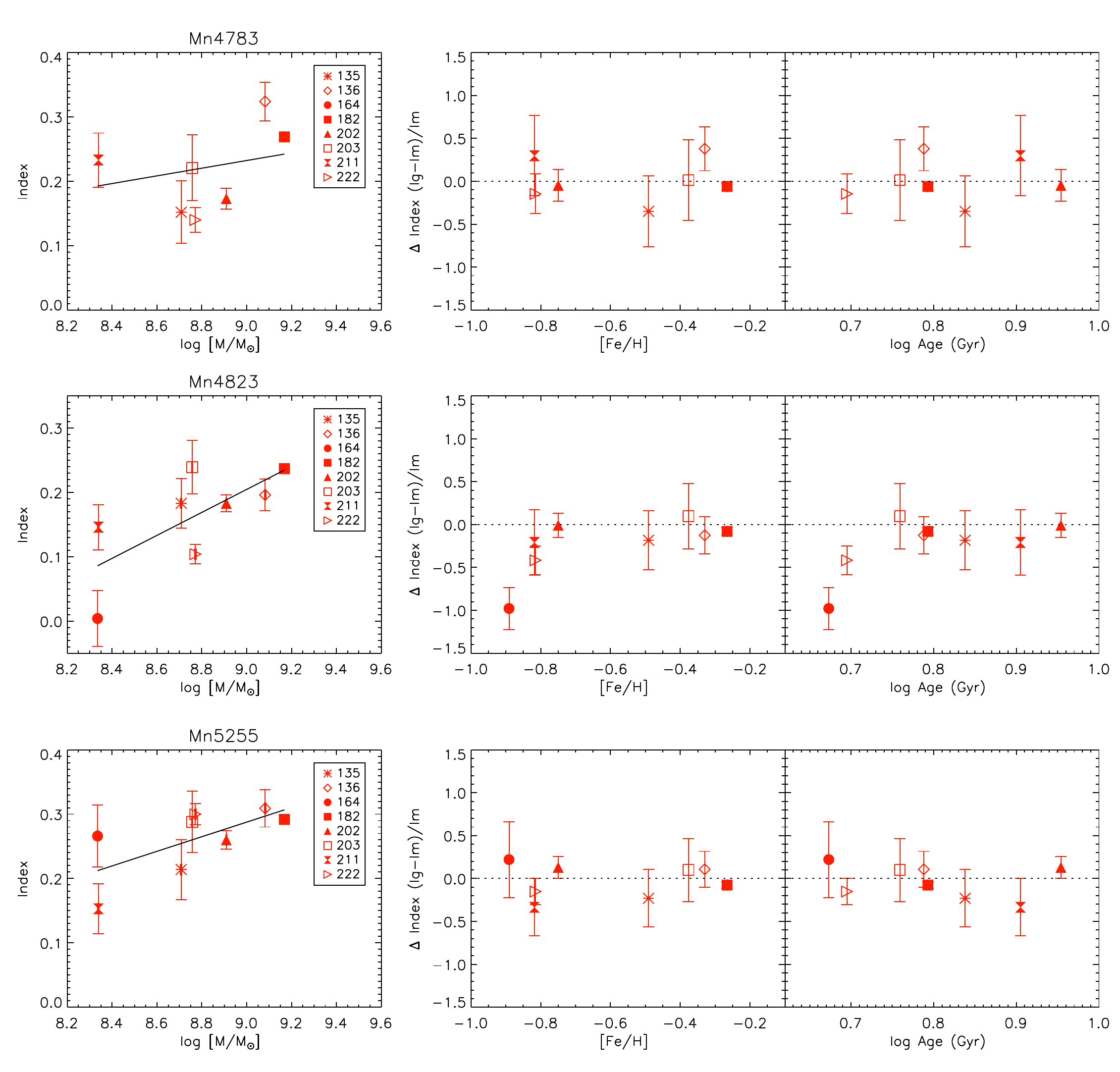}
   \caption{Comparison between index values of PEGASE.HR models and galaxies newly defined Mn lines are plotted as a function of galaxy mass (left panel), are shown in the center and right panels.}
    \label{fig:final_subfig_diff_norm_wg_mn_new}
\end{figure*}

\begin{figure*}
	\includegraphics[width=0.99\textwidth]{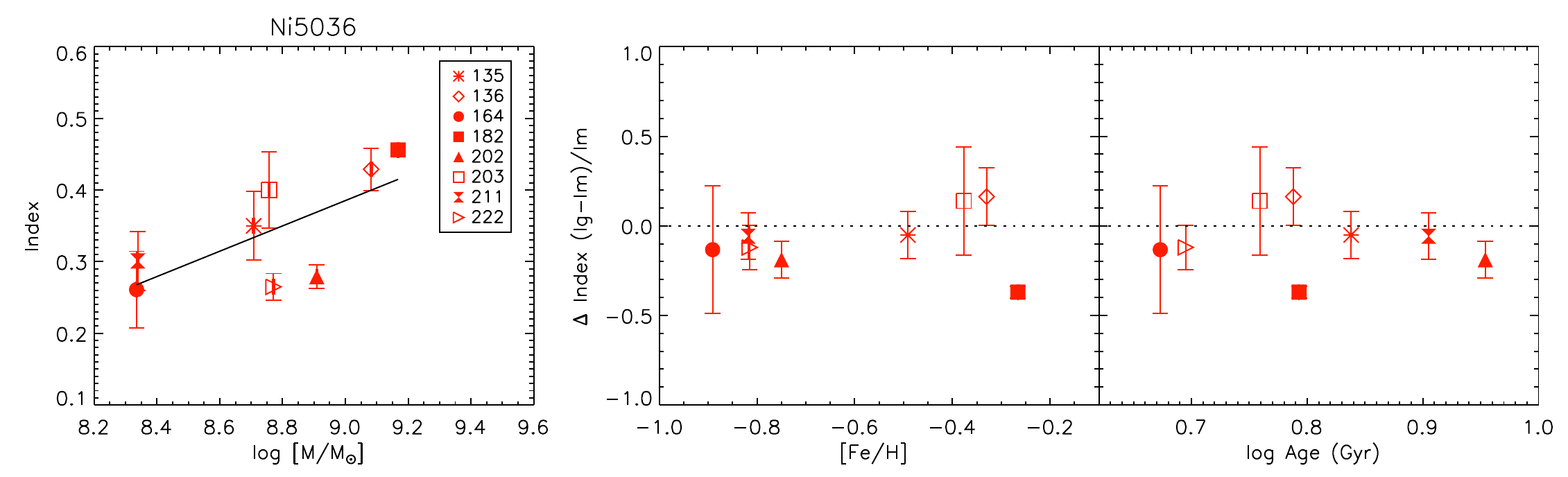}
   \caption{Comparison between index values of PEGASE.HR models and galaxies newly defined Ni line is plotted as a function of galaxy mass (left panel), is shown in the center and right panels.}
    \label{fig:final_subfig_diff_norm_wg_ni_new}
\end{figure*}

\begin{figure*}
	\includegraphics[width=0.99\textwidth]{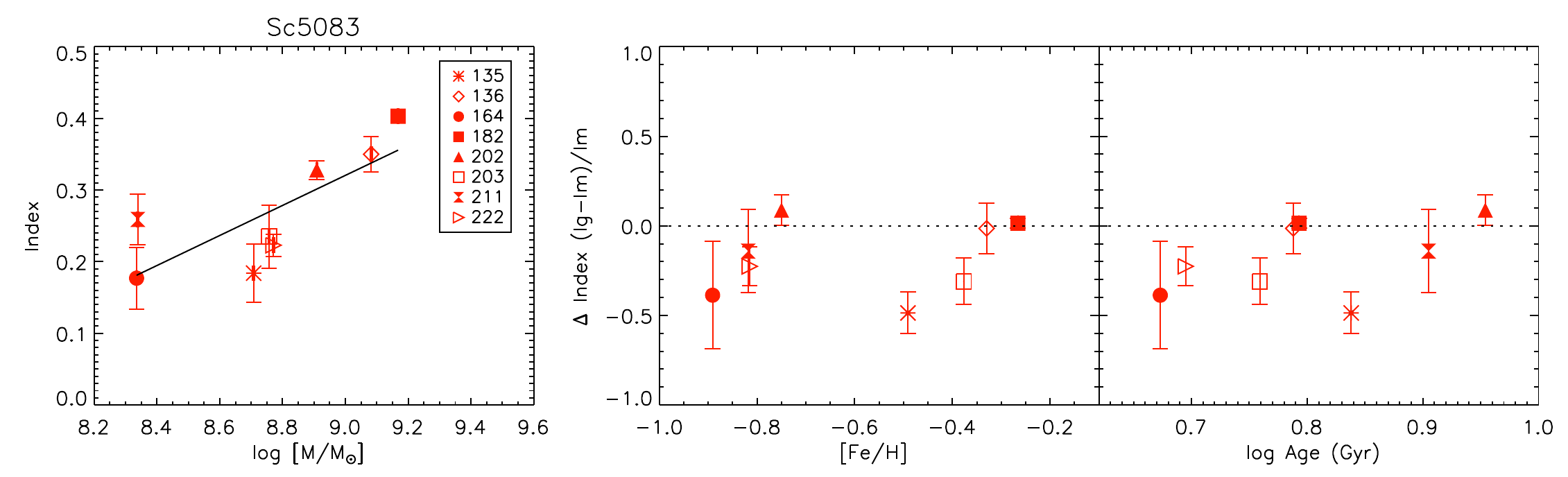}
   \caption{Comparison between index values of PEGASE.HR models and galaxies newly defined Sc line is plotted as a function of galaxy mass (left panel), is shown in the center and right panels.}
    \label{fig:final_subfig_diff_norm_wg_sc_new}
\end{figure*}

\begin{figure*}
	\includegraphics[width=0.99\textwidth]{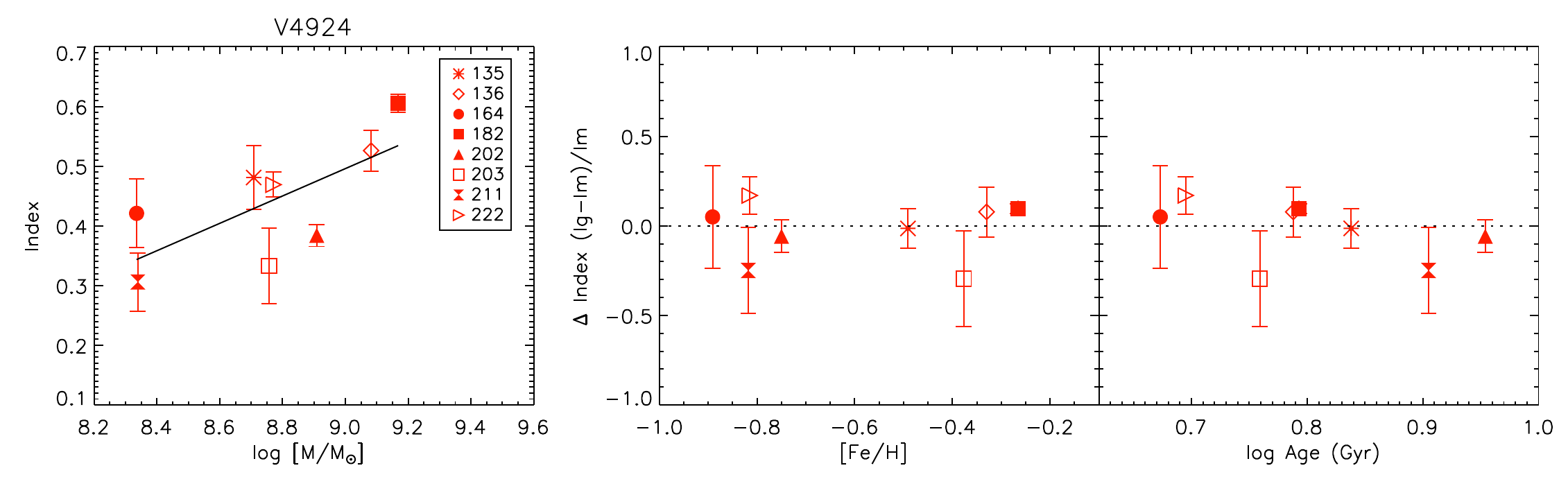}
   \caption{Comparison between index values of PEGASE.HR models and galaxies newly defined V line is plotted as a function of galaxy mass (left panel), is shown in the center and right panels.}
    \label{fig:final_subfig_diff_norm_wg_v_new}
\end{figure*}

\begin{figure*}
	\includegraphics[width=0.99\textwidth]{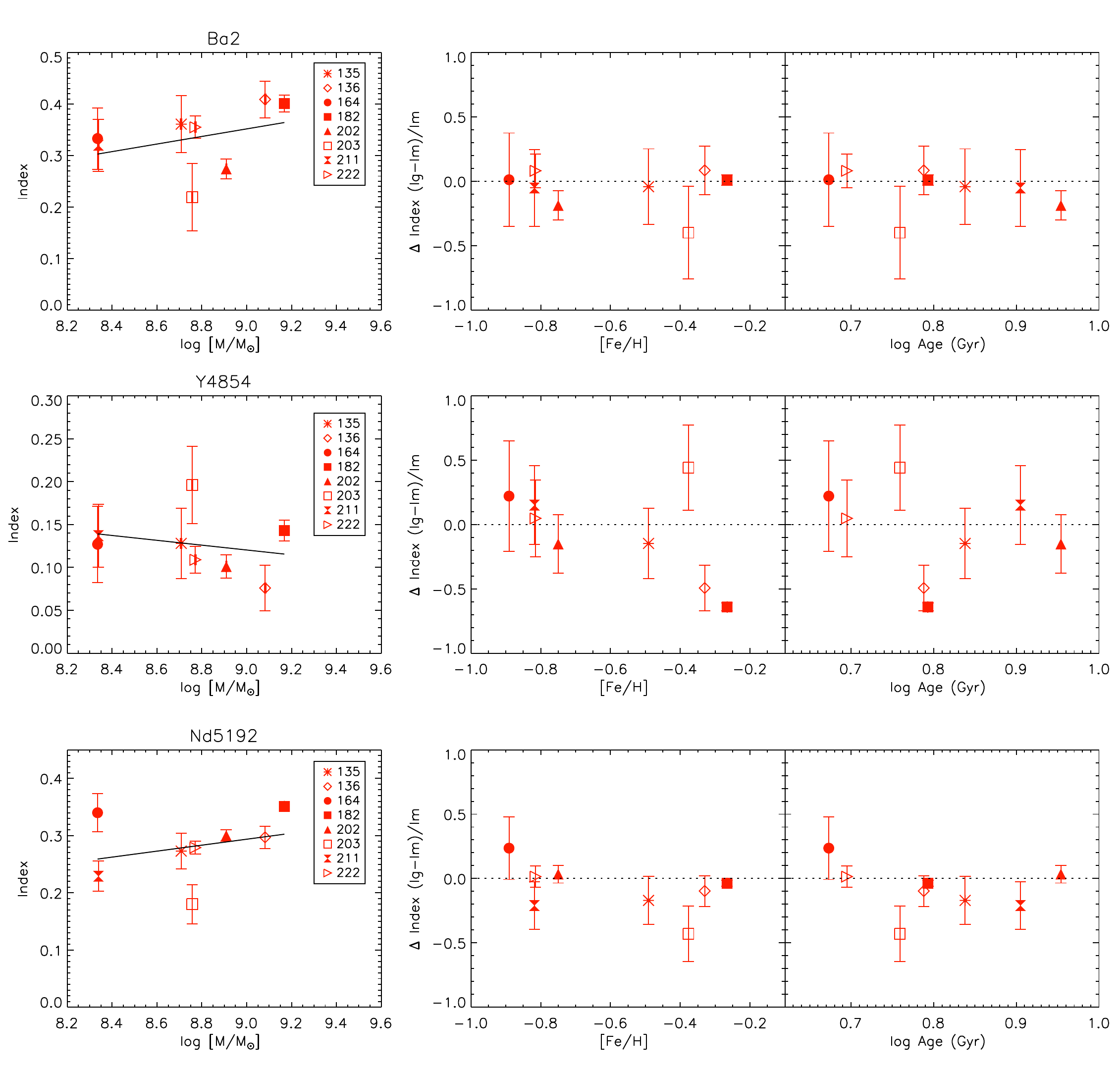}
   \caption{Comparison between index values of PEGASE.HR models and galaxies newly defined heavy elements lines are plotted as a function of galaxy mass (left panel), are shown in the center and right panels.}
    \label{fig:final_subfig_diff_norm_wg_heavy}
\end{figure*}

To test what is the effect of the SSP approach for determining age and metallicity on the line strength measurements, we have done some simulations. For each galaxy we determined two solutions: an SSP-solution (with the PEGASE.HR models) best fiting the observed  ${H}\beta_\text{o}$ and Mg $\textit{b}$ of the galaxy, and a model consisting of a linear combinations of SSPs of Z=0.0004 and $15.9$\,Gyr (PEGASE.HR scale), Z=0.008 and $1$\,Gyr and Z=0.008 and $8$\,Gyr, optimally fitting these 2 absorption lines. Comparing the high resolution indices between these 2 solutions then tells us to what extent the one-SSP approximation is affecting our results. The average differences and standard deviations are given in table \ref{tab:indices_SSPdif}.
The differences are around 10\%. For some lines slightly more, others slightly less. One gets this difference, because of the influence of the young populations, which make the continuum stronger, and therefore all metal lines weaker. Since the results in this paper are qualitative, this simulation shows that our conclusions will not change.

\begin{table*}
 \caption{High-resolution spectral indices: average difference and standard deviation between SSP fit and 3 population fit }
 \label{tab:indices_SSPdif}
 \begin{tabular}{lcclcc} 
Index  &  average($\Delta$(3pop - SSP)/SSP) &  $\sigma$(($\Delta$(3pop - SSP)/SSP)) & Index  &  average($\Delta$(3pop - SSP)/SSP) & $\sigma$(($\Delta$(3pop - SSP)/SSP)) \\
\hline
Ba2	&	-0.169	&	0.074	&	Mn5255	&	-0.112	&	0.096	\\
Ca5041	&	-0.103	&	0.07	&	Ni5036	&	-0.117	&	0.071	\\
Ca5261	&	-0.06	&	0.067	&	Sc5083	&	-0.069	&	0.075	\\
Na4978	&	-0.145	&	0.131	&	V4924	&	-0.049	&	0.09	\\
Ti5014	&	-0.109	&	0.055	&	Y4854	&	-0.165	&	0.133	\\
Ti5064	&	-0.141	&	0.069	&	Nd5192	&	-0.065	&	0.059	\\
Ti5129	&	-0.042	&	0.073	&	Fe4891	&	-0.029	&	0.08	\\
Cr4789	&	-0.102	&	0.09	&	Fe4920	&	-0.035	&	0.087	\\
Cr5072	&	-0.093	&	0.08	&	Fe5226	&	-0.09	&	0.076	\\
Cr5247	&	-0.199	&	0.103	&	Fe5015	&	-0.055	&	0.075	\\
Cr5265	&	-0.031	&	0.079	&	H$\beta_\text{o}$	&	0.05	&	0.071	\\
Cr5275	&	-0.081	&	0.081	&	H$\beta$	&	0.089	&	0.071	\\
Mn4783	&	-0.019	&	0.121	&	Mg $\textit{b}$	&	-0.014	&	0.055	\\
Mn4823	&	0.016	&	0.092	&	~	&	~	&	~	\\

		\hline
  \end{tabular}
 \end{table*}


\section{Results}
In this work we study the abundance ratios of a sample of 8 bright dwarf elliptical galaxies using a new set of 23 high-resolution spectral indices. 
We are interested in studying the differences in the abundance ratios with respect to their more massive galaxy counterparts from a qualitative point of view. However, to increase the reliability of our results, we have selected more than 1 index per element, if possible, to have independent measurements to determine the relative abundance of these elements.

The so-called alpha elements, e.g., O, Mg, Ca, Si and Ti, are predominantly synthesised by alpha capture during the various burning phases in massive stars, and expelled into the ISM by SN II explosions. The abundances of the $\alpha$-elements increase very quickly with time, due to the relatively short main-sequence lifetimes of massive stars. Another group of elements are the Fe-peak elements (Ni, Co, Fe, Mn, etc.) which are mainly produced by SN Ia, whose progenitor lifetimes are much longer. As the evolution of chemical abundances in a given system is closely related to the history of its star formation, abundance ratios such as [$\alpha$/Fe] can be used to explore the SFH for galaxies.

Heavier elements beyond the iron peak, such as Y, Ba and Nd are synthesized by neutron capture where the two important processes occur (the s- and r- processes), followed by $\beta$ decays. These two processes lead to two characteristic abundance patterns. 

For the interpretation of the spectra, we will start discussing the ages and metallicities of the galaxies. We then focus on the relative abundance pattern as derived from the comparison of the observed indices with PEGASE.HR models at higher resolution. We finish this section discussing the behavior of the indices belonging to the various groups of elements. For the indices blueward of $\sim$4800\AA\ we do not include the galaxy FCC 164, since its spectral coverage starts around that wavelengths.

\subsection{Ages and metallicities}

The objects discussed here have a velocity dispersion around 30 kms$^{-1}$, which is similar to the objects with the lowest values in \citet{sybliskaetal.2017MNRAS}. The mean age and metallicity are $10.6$\,Gyr, and [M/H]=-0.43, respectively. The galaxies appear to be slightly older and more metal rich than those of \citet{sybliskaetal.2017MNRAS}, although their sample mostly consists of more massive galaxies. A sample that offers a better comparison is that of \citet{tolobaetal.2014ApJS..215...17T}, for which all velocity dispersions lie between 20 and 40 kms$^{-1}$, as is the case in this paper. These authors find a mean age of $6.1$\,Gyr, and mean metallicity of -0.61, also slightly younger and more metal poor in comparison to our sample. These differences could be due the fact that both the galaxies of Toloba et al. and Sybilska et al. are in Virgo, while these are in Fornax, a cluster which is more evolved in the central regions. In fact the study of \citet{hamrazetal.2019A&A...625A..94H} shows that the scatter in the $g-z$ colour in the mass range of the dwarfs discussed here is considerably larger in Virgo than in Fornax, indicating more young stellar populations and more dust in Virgo. Beside that  \citet{kuntschneretal.2000MNRAS.315..184K} shows that all their ellipticals in Fornax cluster were consistently old, while the S0 galaxies show lower ages. However it is still possible that the result that the Virgo galaxies are less metal poor might be attributed in part to the age-metallicity degeneracy. In this case if the Virgo galaxies had the same metallicity as the ones in Fornax, their ages would have to be younger, making the difference in ages even larger. When comparing the metallicities obtained in this work with the compilation of \citet{caldwell.2003AJ....125.2891C} we find lower values.  Their ages are also lower, so it is possible that the age-metallicity degeneracy is at work, partly due to the fact that the employed set of indices to determine these parameters are not free from the abundance ratio effects.

\subsection{Abundance ratios of the stellar population models}

The stellar population models used in this work, i.e. PEGASE.HR and MILES, are based on observed Galactic stars, mostly obtained with intermediate-size telescopes, which means that  their spectra are imprinted with the abundance ratio trends of The Galaxy. In particular the \'Elodie library (\citealp{prugnielandsoubiran.2001A&A...369.1048P}) with resolution 10000, which feed the PEGASE.HR models, is composed of bright stars in the Galactic disk, although it also includes some stars at lower metallicity in the thick disk and halo. Therefore the SSP spectra computed with PEGASE.HR models also follow this pattern. Similarly, the lower resolution MILES SSP model spectra also follow the same pattern, as inferred from the abundance determinations of \citet{miloneetal.2011MNRAS.414.1227M} and \citet{garciaperezetal.2021MNRAS.tmp..304G}. Therefore the base models employed here have [Mg/Fe] = 0 at [Fe/H]=0, [Mg/Fe] $\sim$ 0.2 for [Fe/H]=-0.4 and $\sim$ 0.4 for [Fe/H]=-1.0. This has to be taken into account for discussing the abundance ratio determinations in dwarf elliptical galaxies as already shown by \citet{sybilskaetal.2018MNRAS.476.4501S}.

\subsection{The $\alpha$-elements: Ca, Mg and Ti }

\textit{Calcium:} The two Calcium lines that are present in the covered spectral range are centered at 5041 \AA\ and 5261 \AA. We see that the measured index strengths increase with galaxy mass, being more pronounced for the line at 5041 \AA. The observed index values are very close, although slightly weaker, to the ones expected from the models. Fig. \ref{fig:final_subfig_diff_norm_wg_ca_new} shows the ratio between the observed galaxy and the ELODIE models, which were specifically selected to have the same age and metallicity as the observed galaxy. Such a difference can be attributed in part to the fact that the base models have a slightly enhanced abundance ratio of Ca, i.e. another $\alpha$ element, in the sub-solar metallicity regime that is characteristic of these galaxies. Therefore the relatively lower line-strengths shown by our galaxies indicate that their Ca abundance ratio is less enhanced than in the reference models and, therefore, closer to the solar-scaled value. Note that we are not attempting at matching the observed line-strengths as we are only interested in performing a qualitative assessment of the abundance ratios in these galaxies.

\textit{Magnesium:} This $\alpha$-element, which is produced by massive stars during the hydrostatic He burning phase, is easily measured from the Mg $\textit{b}$ Lick index in the observed wavelength range. Note that this is the only strong Mg line in this spectral range. We present the comparison between observed and model line-strengths in Fig. \ref{fig:final_subfig_diff_norm_wg_mg_new}. Remarkably the Mg $\textit{b}$ shows a very strong relation with galaxy mass even at these low masses, and in such a small mass window. We find that the observed index values are slightly lower than in the models, which point to an abundance ratio that is closer to the
Milky-Way abundance ratio at this subsolar metallicity regime. This result is very similar to that obtained for the Ca lines, in quite good agreement with \citet{Sen.2018MNRAS.475.3453S}.  

\textit{Titanium:} The wavelength region observed with SAMI includes three of the stronger Ti lines.  The measured index strengths are compared to the reference models in Fig. \ref{fig:final_subfig_diff_norm_wg_ti_new}. We see that the three lines increase very strongly with galaxy mass as it was the case for the Mg $\textit{b}$ index. We see also that the three indices show similar relative deviations with respect to the models, which can be taken as a good indication of their sensitivity to the Ti abundance ratio. Ca, and Ti are mostly produced during SN II explosions, and in general they trace one another. For these galaxies we find that the Ti line-strengths are lower than the model predictions. Note that the deviations are slightly larger for the Ti5129 index, probably because this line is significantly fainter (the index values are around 0.15\AA) than the other two, implying that systematic errors might have a larger effect than represented by the errorbars.

\subsection{The Light odd-Z element: Na}
\textit{Sodium:} Na can be measured using the strong absorption features of the NaD doublet at 5890 and 5896 \AA. However only the Na line at 4978 \AA\ is available in the observed wavelength range. The index strengths observed for these dEs are lower than those predicted by the models with similar age and metallicity as shown in  Fig. \ref{fig:final_subfig_diff_norm_wg_na_new}. The obtained deviation is significantly larger than those shown by the Ca, Mg and Ti lines. We see also that Na4978 shows a clear trend with [Fe/H] and age. All the results inferred for these galaxies, which are based on this line, are in good agreement with those presented in \citep{Sen.2018MNRAS.475.3453S}.

\subsection{Fe-peak elements: Cr, Mn, Ni, Sc, V and Fe }
Iron peak elements are the result of a complex nucleosynthesis process. Fe and Ni are produced mainly by complete explosive Si burning in the deepest layers, 
while Cr, Mn and V are produced mainly in the outer incomplete Si-burning layers. Sc is synthesized during explosive Oxygen / Neon burning and can be considered a transition element, intermediate between the $\alpha$-elements and the iron-peak elements.

\textit{Chromium:} Cr lines are represented with five lines in the observed wavelength range. All Cr lines correlate well with galaxy mass, as shown in Fig. \ref{fig:final_subfig_diff_norm_wg_cr1}. The comparison between the observed and model index strengths show that the galaxy measurements are slightly below 0 in these plots, with the largest deviations being obtained for the Cr5072. As Cr belongs to the Fe peak elements the observed deviations suggest an abundance ratio slightly below the solar-scale value.

\textit{Manganese:} For Mn we use three line indices, which are shown in Fig. \ref{fig:final_subfig_diff_norm_wg_mn_new}, in the observed wavelength range. Although faint all the three lines suggest that an abundance ratio slightly below solar, similarly to Cr. 

\textit{Nickel:} The abundance of Ni is studied by one line at 5036 \AA. Fig. \ref{fig:final_subfig_diff_norm_wg_ni_new} shows that the measurements of the galaxies are lower than the values predicted by the models, leading to the same result obtained for Cr and Mn. 

\textit{Scandium:} Scandium is represented by one line at 5083 \AA, which shows very good correlation with mass. The observed index values lie slightly lower below 0 in Fig. \ref{fig:final_subfig_diff_norm_wg_sc_new}, similarly to Cr, Mn and Ni.

\textit{Vanadium:} We have one, relatively strong, V line at 4924 \AA, which suggest an abundance ratio rather close to solar as shown in Fig.~ \ref{fig:final_subfig_diff_norm_wg_v_new}.

\subsection{Neutron capture elements: Ba, Y, Nd}
Heavy elements are those with atomic number higher than 30, like Yttrium (Y), Barium (Ba) and Neodymium (Nd). They can only be produced by neutron capture elements that are exposed to high neutron flux. Iron peak elements are the most efficient seeds to capture neutrons to create heavier elements. There are two main paths to form these elements: the s-process (or slow process) and the r-process (rapid process).

The s-process occurs when the neutron flux is not very high, so that the intervals between neutron captures are long compared to the beta decay characteristic timescale of an unstable nucleus.

\citet{tolstoyetal.2009ARA&A..47..371T} compare Ba, Y and Eu abundances in Local Group dwarf Spheroidals and in the Milky Way. In dSphs, the early evolution of all neutron-capture elements is dominated by the r-process, after which the s-process starts taking over from the r-process. The metallicity of switch from r- to s-process is the same as the [$\alpha$/Fe] knee ([Fe/H] $\sim$ -1.8), i.e. much lower than in the Milky Way. In the latter, Ba and Y are dominated by the r-process for [Fe/H] $\leq$ -2.0, while the s-process dominates at higher metallicities, for instance more than 80$\%$ of the solar Ba is originating from the s-process.

\textit{Barium:} is represented by one line, Ba2, at 4895 \AA. \textit{Yttrium:} is well represented by a line at 4854 \AA\ in our dwarf galaxies. \textit{Neodymium:} can be measured by a line centered at 5192 \AA. All heavy elements are found to have abundances slightly lower than the solar-scale value as shown in Fig \ref{fig:final_subfig_diff_norm_wg_heavy}.

\section{DISCUSSION}

\begin{figure*}
	\includegraphics[width=0.99\textwidth]{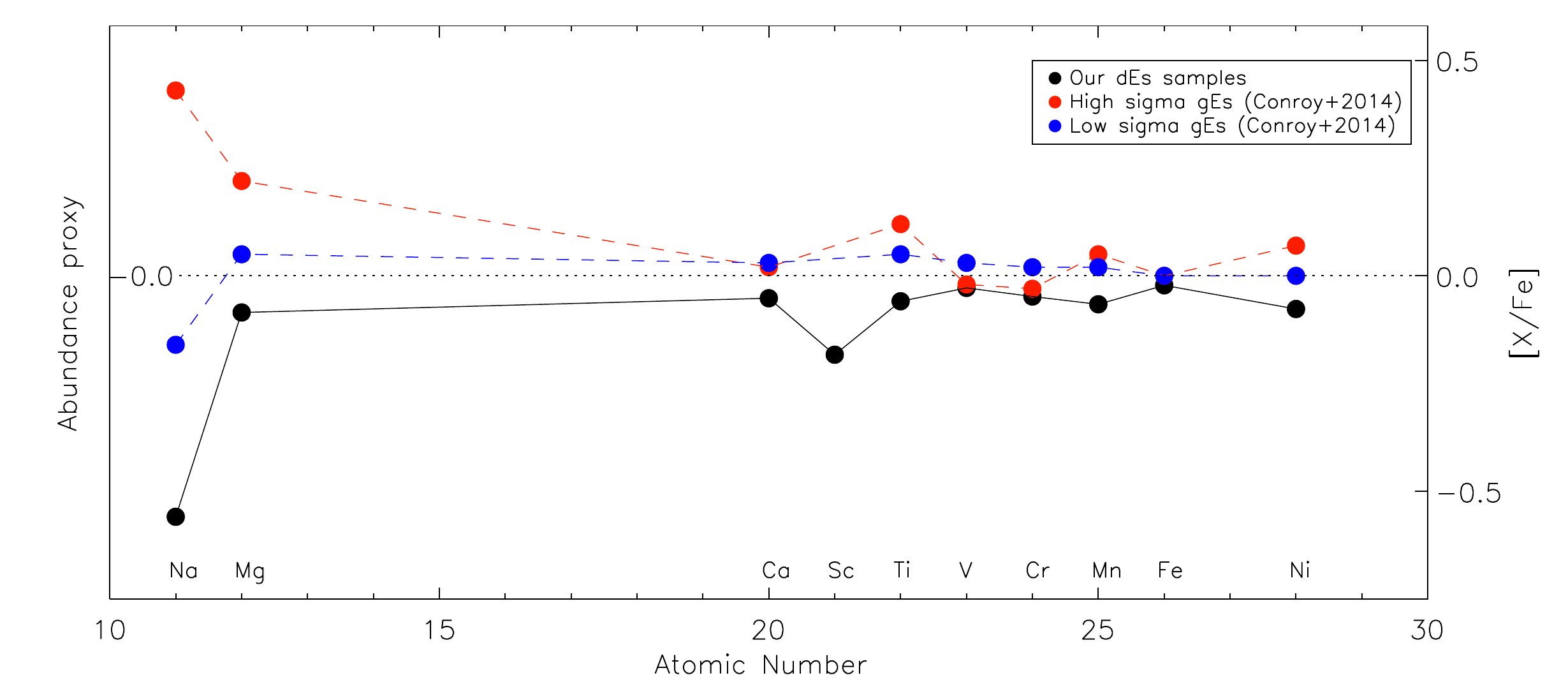}
   \caption{Summary of the abundance trends derived here from our work and low- and high- velocity dispersion gEs (Conroy et al. 2014). Note that we indicate the abundance ratio values in the right vertical axis, which apply to the two gEs samples shown here. The abundance proxy estimates for our dwarf galaxies are reflected in the left vertical axis, where we indicate only the relative non-calibrated index deviations with respect to the reference base SSP models, represented by the 0.0 value.}
    \label{fig:elements_average_all}
\end{figure*}

The main results from this paper are summarized in Fig. \ref{fig:elements_average_all}, where we show the proxy for the abundance ratios for the 11 elements derived here. For comparison we also provide the "true" abundance ratio estimates for more massive ellipticals with very different velocity dispersion values by \citet{conroyetal.2014ApJ...780...33C}. We note that the derived abundance ratio proxies for our galaxies as relative index deviations with respect to the reference base SSP models. These models are fed with empirical stars and therefore follow the Milky-Way abundance pattern, which is represented in Fig.~\ref{fig:elements_average_all} by a "0.0" value in its left vertical axis. It is worth recalling that this reference valued does not mean solar-scaled abundance (which is only true at solar metallicity). We do not calibrate these deviations in terms of abundance element ratios. In contrast the right vertical axis indicates the abundance ratio values for the gEs shown in this figure. However, having these two estimates in the same plot allow us to compare visually these trends. It is worth recalling that such comparison is provided with the only purpose of assessing the difference in the derived abundance trends between these three families of galaxies, rather than an in-depth quantitative analysis of the their true abundances.

For our galaxies we find that the abundance ratio proxies of all elements are either below or close to the solar-scale value. This means that these dwarf galaxies form their stars slowly, like in the disk of the Milky Way. The results obtained here for Mg and Ca are in rather good agreement with the ones published in \citet{Sen.2018MNRAS.475.3453S}. This behavior fits nicely on a trend with increasing mass, from dwarfs, through Milky-way type galaxies, to the most massive giant galaxies, as analysed by \citet{conroyetal.2014ApJ...780...33C}. Galaxies with $\sigma$ $\sim$ 100 kms$^{-1}$  have abundance ratios that are close to solar, while in massive galaxies the $\alpha$-elements are enhanced, together with a few other elements (see \citealp{conroyetal.2014ApJ...780...33C}). In particular, Na shows a different relation, being significantly lower than solar for dwarfs as compared to the solar-scale value, and much higher for giant ellipticals (e.g. \citealp{smithetal.2015MNRAS.454.2502S, labarberaetal.2017MNRAS.464.3597L}).

We can make a few remarks for some specific elements. We find that [Mn/Fe] seems to be a bit lower than abundance ratios of other Fe-peak elements. This result is likely consistent with the [Mn/Fe] ratios derived for Galactic stars in the sub-solar metallicity regime of our dEs, i.e.   [Mn/Fe]$\sim$-0.2 around [Fe/H]$\sim$-0.5
\citealp{gratton.1989A&A...208..171G,mcwilliametal.2003ApJ...592L..21M}). The latter reference also shows that in the Galactic bulge and in the Sagittarius dwarf the [Mn/Fe] abundance is once again about 0.2 lower than the solar-scale value. Although the abundance of V has not yet been well determined, it is still thought that [V/Fe] is around  $\sim$ 0.0 at all metallicities \citep{grattonandsneden.1991A&A...241..501G}. If the Mn deficiencies are due to a neutron excess dependence, then V and Sc are also expected to follow the same trend, as is inferred for the present sample of dEs, although we only rely on a single line in this study.

Na is a very interesting element that is produced in the interiors of massive stars depending on  the neutron  excess.  This  means that it depends on the initial heavy element abundance in the star. We find that Na line strengths show low values compared to the models in the dEs. These results are in good agreement with the results of \citet{Sen.2018MNRAS.475.3453S}, which show that for Virgo dEs the [Na/Fe] is under-abundant with respect to the solar-scale value. Note that this is in sharp contrast with the abundance found for giant ellipticals, which show strongly overabundant [Na/Fe] abundance ratios (e.g. \citealp{conroyetal.2014ApJ...780...33C,labarberaetal.2017MNRAS.464.3597L}). This abundance ratio also shows a strong correlation with galaxy mass.

The similarities with disk-like abundance ratios may suggest that dEs originate from star forming galaxies with rather slow formation. Given that dEs are mostly found in dense environments (e.g. \citealp{binggelietal.1988ARA&A..26..509B}), where star forming galaxies are scarce, point to late types spirals or star forming dwarfs (e.g. dwarf irregulars), losing their interstellar medium when entering the cluster through ram pressure stripping (see e.g. \citealp{boselliandgavazzi_2014A&ARv..22...74B,Choque-Challapaetal.2019MNRAS.490.3654C}). It is in principle also possible that galaxy interactions, i.e., galaxy harassment \citep{mooreetal.1996Natur.379..613M} causes this transformation, but in such a case it would be expected considerable changes in the properties of dEs as a function of position in the cluster, which have not been seen up to now (e.g. \citealp{venholaetal.2019A&A...625A.143V}). The scenario of star forming dwarfs transforming into quiescent dEs agrees with recent kinematic results of \citet{scottetal.2020MNRAS.497.1571S}. These authors show that the rotational support in dEs is similar or slightly lower than in star forming dwarfs. No strong relation was found between the rotational support and the cluster-centric distance. If dEs form from late-type spirals, such as Sd or Sm galaxies, a considerable amount of angular momentum has to be lost, which points to galaxy harassment as the process that is responsible for this transformation. 

We will discuss the star formation histories of our present sample of galaxies and other dEs in an upcoming paper (Romero et al., in preparation). The results obtained here indicate long star formation time scales for these galaxies.


\section{Conclusions} 
\begin{itemize}

\item This is the first time that an attempt has been made to determine abundance ratio proxies of many elements in unresolved dwarf galaxies outside the Local Group. It was done using a sample of dE galaxies, observed using SAMI at the AAT at a resolution of $\sim$ 5000. These abundance proxies must be understood as relative index deviations with respect to the model predictions that follow the Milky-Way abundance pattern.

\item We analysed the absorption line-strengths of our new high resolution system of indices for a sample of 8 Fornax dwarf galaxies, which had been classified as cluster members in \citet{venholaetal.2019A&A...625A.143V}.

\item Taking advantage of the high resolution spectral data and low velocity dispersion (40 kms$^{-1}$) of these galaxies we were able to measure the new high resolution indices with our SAMI data and compare the obtained index values to the predictions of the PEGASE.HR stellar population models.
We derive abundance proxies for 11 elements and compare them with the abundance ratios of massive galaxies by \citet{conroyetal.2014ApJ...780...33C} as a function of mass as traced by their velocity dispersion. We find that the dwarf galaxies in our sample have abundance ratios that are close to solar-scale or slightly below this value. This pattern is consistent with an extrapolation of the abundances of massive galaxies  \citep{conroyetal.2014ApJ...780...33C} to lower masses.

\item We find that these dwarf galaxies have [Na/Fe] abundance ratios that are considerably lower than the solar-scale value. This result is in good agreement with Virgo dEs, which show that [Na/Fe] is under-abundant with respect to the solar element partition \citep{Sen.2018MNRAS.475.3453S} . 

\item This work indicates the large potential for future studies of low mass stellar systems, with powerful instruments, such as X-Shooter on the VLT.

\end{itemize}
\section*{Acknowledgements}
\c{S}. \c{S}. (Aydemir) acknowledges support from grant 2218-National Postdoctoral Research Fellowship Program for Turkish Citizens under project number 118C553 and application number 1929B012100279 from the The Scientific and Technological Research Council of Turkey (TUBITAK). RFP acknowledges financial support from the European  Union’s  Horizon  2020  research  and  innovation program  under  the  Marie  Sklodowska-Curie  grant  agreement No. 721463 to the SUNDIAL ITN network. A.V. acknowledges support from grants PID2019-107427GB-C32 from the Spanish Ministry of Science, Innovation and Universities (MCIU).

\section*{Data Availability}

The raw data used in this paper, from the SAMI instrument, are available from the AAT data archive. The reduced data are cubes are available upon request, and will be made available in the near future together with a future paper on the star formation histories of dwarf galaxies in the Fornax Cluster.





\bibliographystyle{mnras}
\bibliography{manuscript} 





\appendix
\section{Table}

\begin{table}
	\centering
	\caption{INDEX DEFINITIONS.}
	\label{tab:indexdefinitions}
	\begin{tabular}[h!]{ lccc } 
		\hline
		Name & Index Bandpass & Blue Continuum & Red Continuum  \\

		  & \AA & \AA & \AA \\
		\hline
Ba2	    &	4930.000--4935.800	&	4894.400--4896.800	&	4948.600--4951.800	   \\
Ca5041	&	5036.200--5043.740	&	5032.720--5034.790	&	5056.820--5059.460	   \\
Ca5261	&	5259.925--5263.260	&	5257.400--5259.500	&	5277.310--5279.410	   \\
Cr4789	&	4787.350--4791.120	&	4768.880--4771.080	&	4793.450--4795.700	   \\
Cr5072	&	5071.095--5073.570	&	5056.820--5059.460	&	5092.920--5095.800	   \\
Cr5247	&	5244.700--5248.495	&	5236.250--5238.250	&	5256.200--5259.200	   \\
Cr5265	&	5263.700--5268.340	&	5257.400--5259.500	&	5277.310--5279.410	   \\
Cr5275	&	5273.700--5277.600	&	5257.400--5259.500	&	5277.310--5279.410	   \\
Fe4891	&	4889.125--4893.040	&	4856.250--4858.250	&	4893.800--4896.200	   \\
Fe4920	&	4914.425--4921.850	&	4894.400--4896.800	&	4948.600--4951.800	   \\
Fe5226	&	5219.800--5230.240	&	5212.400--5214.500	&	5236.250--5238.250	   \\
Mn4783	&	4781.500--4785.560	&	4768.880--4771.080	&	4793.450--4795.700	   \\
Mn4823	&	4822.420--4825.780	&	4814.600--4820.200	&	4844.500--4846.900	   \\
Mn5255	&	5253.200--5257.260	&	5236.250--5238.250	&	5256.200--5259.200	   \\
Na4978	&	4976.300--4979.780	&	4948.730--4952.190	&	4986.600--4988.760	   \\
Nd5192	&	5190.940--5193.530	&	5174.500--5178.500	&	5198.540--5201.760	   \\
Ni5036	&	5034.440--5037.420	&	5032.720--5034.790	&	5056.820--5059.460	   \\
Sc5083	&	5081.150--5084.920	&	5056.820--5059.460	&	5092.920--5095.800	   \\
Ti5014	&	5012.350--5016.120	&	5002.400--5005.100	&	5032.260--5034.330	   \\
Ti5064	&	5063.380--5067.280	&	5056.820--5059.460	&	5092.920--5095.800	   \\
Ti5129	&	5127.800--5130.680	&	5118.800--5121.000	&	5143.710--5147.140	   \\
V4924	&	4922.000--4927.480	&	4894.400--4896.800	&	4948.600--4951.800	   \\
Y4854	&	4853.680--4856.410	&	4844.800--4846.900	&	4856.500--4858.500	   \\
	
		\hline
	\end{tabular}
\end{table}

\begin{table}
\centering
\caption{Age-metallicity dependence.  Column 1: index name. Columns 2 and 3: metallicity and age dependence. Columns 4: indicating age or metallicity indicator. The elements H and Mg are presented here as a reference elements; a= age indicator, m=metallicity indicator, a*=more affected by age, m*=more affected by metallicity, blank ones are affected by both.}
	\label{tab:age_metallicity_dependence}
	\begin{tabular}[h!]{ lccc } 
		\hline
	Index & A$_i$  & B$_i$ & Indicator \\
		\hline
		Ba2	&	0.362	$\pm	0.138	$	&	0.532	$\pm	0.102	$	&	a*	\\
        Ca5041	&	0.399	$\pm	0.123	$	&	0.472	$\pm	0.123	$	&	…	\\
        Ca5261	&	0.414	$\pm	0.089	$	&	0.473	$\pm	0.089	$	&	…	\\
        Cr4789	&	0.502	$\pm	0.169	$	&	0.530	$\pm	0.193	$	&	…	\\
        Cr5072	&	0.460	$\pm	0.130	$	&	0.504	$\pm	0.143	$	&	…	\\
        Cr5247	&	0.928	$\pm	0.111	$	&	1.824	$\pm	1.019	$	&	…	\\
        Cr5265	&	0.443	$\pm	0.124	$	&	0.485	$\pm	0.132	$	&	…	\\
        Cr5275	&	0.411	$\pm	0.127	$	&	0.347	$\pm	0.142	$	&	m*	\\
        Fe4891	&	0.552	$\pm	0.238	$	&	0.747	$\pm	0.164	$	&	m*	\\
        Fe4920	&	0.475	$\pm	0.179	$	&	0.507	$\pm	0.180	$	&	…	\\
        Fe5226	&	0.482	$\pm	0.097	$	&	0.561	$\pm	0.100	$	&	…	\\
        Mn4783	&	0.715	$\pm	0.166	$	&	0.719	$\pm	0.273	$	&	…	\\
        Mn4823	&	0.491	$\pm	0.183	$	&	0.506	$\pm	0.160	$	&	…	\\
        Mn5255	&	0.556	$\pm	0.135	$	&	0.500	$\pm	0.178	$	&	…	\\
        Na4978	&	0.675	$\pm	0.151	$	&	0.767	$\pm	0.338	$	&	…	\\
        Nd5192	&	0.395	$\pm	0.111	$	&	0.501	$\pm	0.103	$	&	m*	\\
        Ni5036	&	0.417	$\pm	0.122	$	&	0.466	$\pm	0.132	$	&	…	\\
        Sc5083	&	0.435	$\pm	0.127	$	&	0.560	$\pm	0.118	$	&	a*	\\
        Ti5014	&	0.385	$\pm	0.161	$	&	0.567	$\pm	0.133	$	&	a*	\\
        Ti5064	&	0.442	$\pm	0.088	$	&	0.648	$\pm	0.080	$	&	a*	\\
        V4924	&	0.425	$\pm	0.179	$	&	0.382	$\pm	0.176	$	&	m*	\\
        Y4854	&	0.918	$\pm	0.474	$	&	0.908	$\pm	0.401	$	&	…	\\
        Mg {\textit{b}} 	&	0.760	$\pm	0.192	$	&	0.438	$\pm	0.161	$	&	m	\\
        H{\textit{$\beta$}}	&	0.163	$\pm	0.075	$	&	0.557	$\pm	0.050	$	&	a	\\
        H\textit{{$\beta_o$}}	&	0.061	$\pm	0.032	$	&	0.456	$\pm	0.019	$	&	a	\\
	\hline
	\end{tabular}
\end{table}

\begin{table}
	\centering
	\caption{Abundance ratio proxies for a number of elements.}
	\label{tab:abundance_proxy}
	\begin{tabular}[h!]{ lc } 
		\hline
		Name & Abundance proxy  \\
		\hline
        Na	&	-0.6	   \\
        Mg	&	-0.1	   \\
        Ca	&	-0.1	   \\
        Sc	&	-0.2	   \\
        Ti	&	-0.1	   \\
        V	&	0.0	   \\
        Cr	&	0.0	   \\
        Mn	&	-0.1	   \\
        Fe	&	0.0	   \\
        Ni	&	-0.1	   \\
	    \hline
	\end{tabular}
\end{table}

\newpage
\section{All Figures for the behaviour of the indices as a function of metallicity, age and spectral resolution}
\label{sec:apB}
\begin{figure*}
	\includegraphics[width=0.99\textwidth]{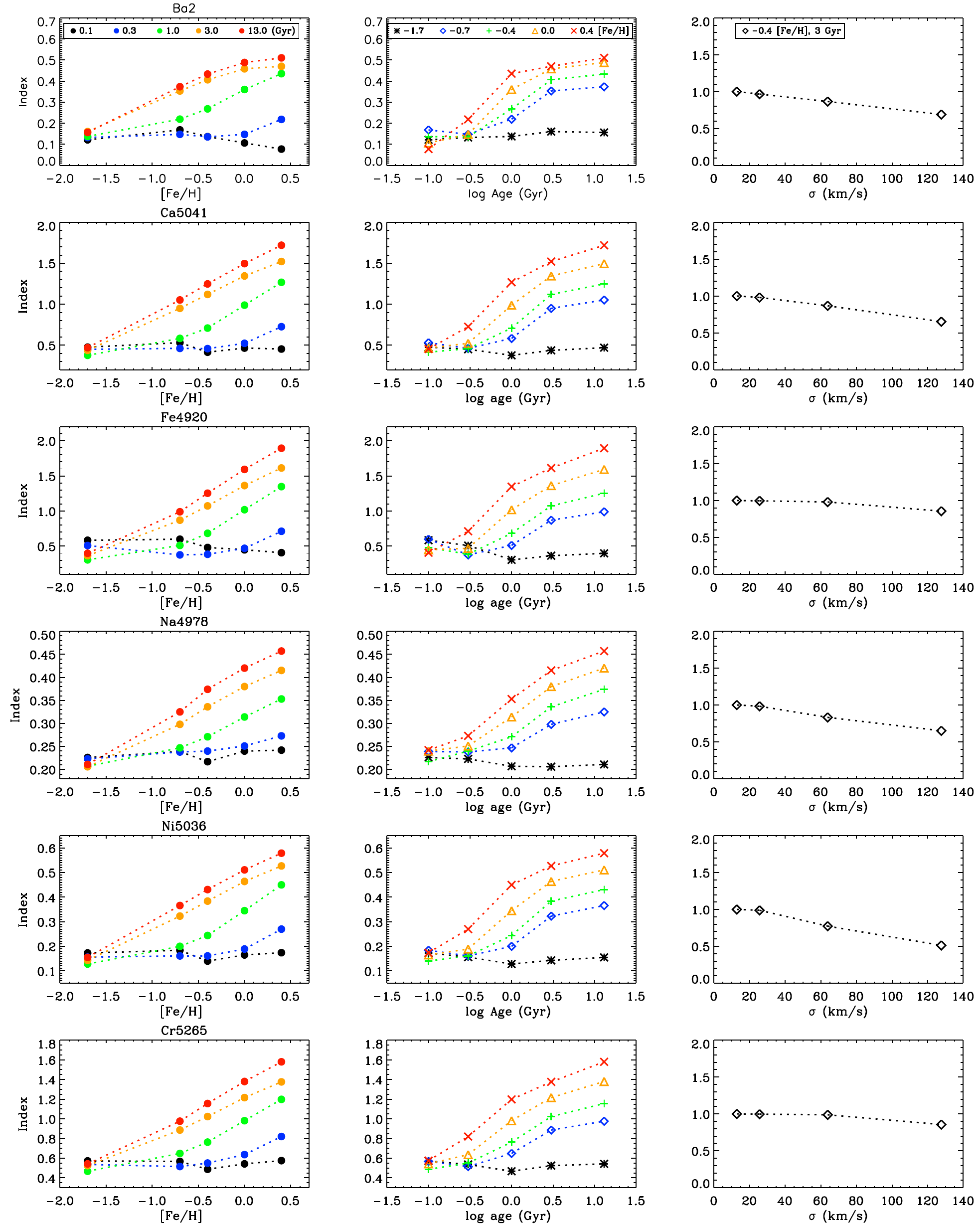}
    \caption{Behaviour of the indices as a function of metallicity, age and spectral resolution. On the left is shown the indices vs. metallicity ([Fe/H]= -1.7, -0.7, -0.4, 0.0, 0.4); each age is shown as a circle with different colour. In the middle is shown the behaviour of indices vs. age (0.1, 0.3, 1, 3, 13 Gyr); each metallicity has a different symbol with different colour. The left and middle panels have been calculated for $\sigma$= 25 km\,s$^{-1}$. On the right is shown the behaviour of indices as a function of velocity dispersion (with $\sigma$= 10-130 km\,s$^{-1}$). Their dependence on $\sigma$ are given for the case of a model with t = 3 Gyr and [Fe/H]= -0.4, typical for the dwarf galaxies discussed in section 5.1. Here 6 example indices are shown.}
    \label{fig:final_subfig_v3_2022}
\end{figure*}

\begin{figure*}
	\includegraphics[width=0.99\textwidth]{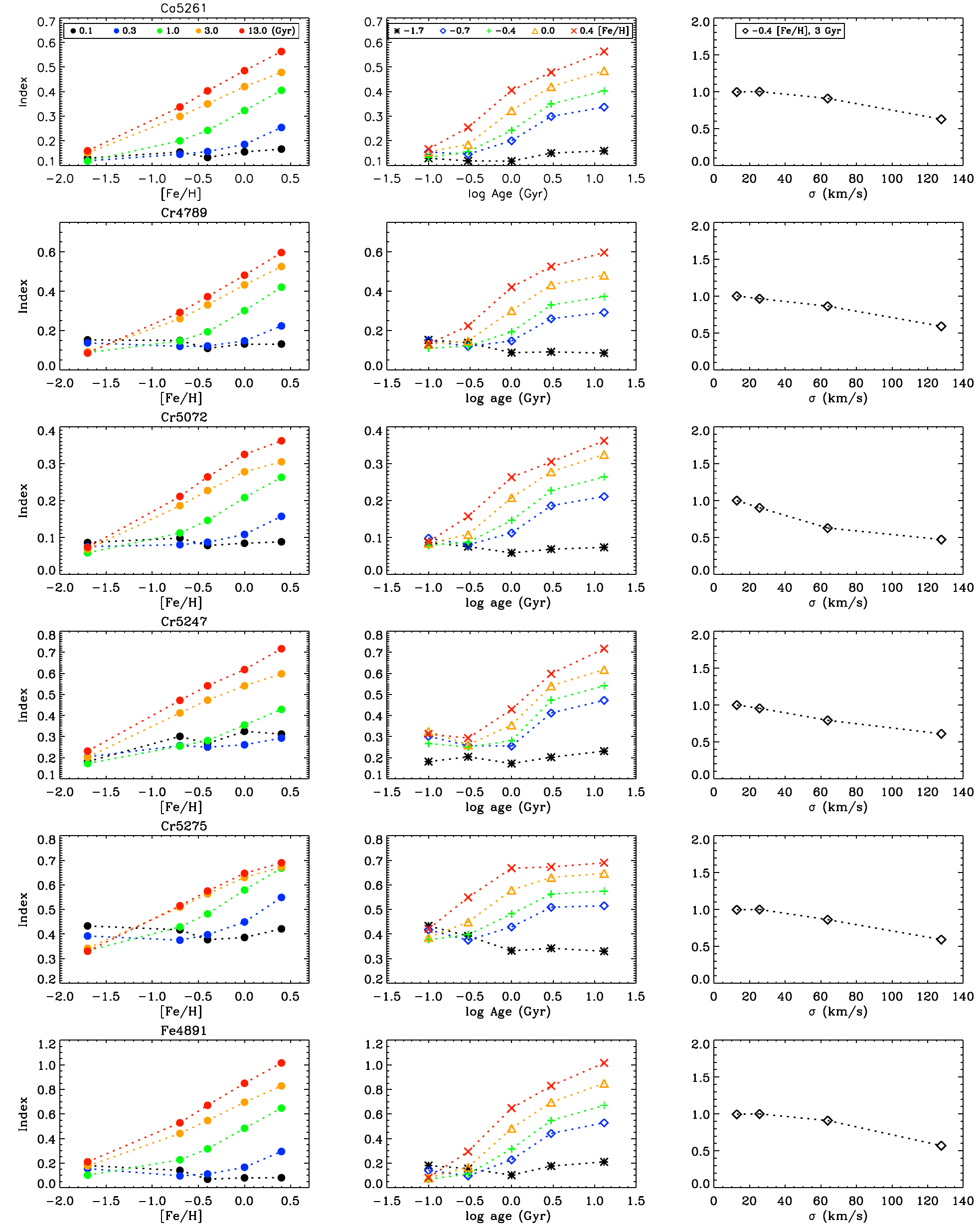}
    \caption{Same as Figure~\ref{fig:final_subfig_v3_2022}, now for the indices: Ca5261, Cr4789, Cr5072, Cr5247, Cr5275 and Fe4891.}
    \label{fig:final_subfig_v3_1_2022}
\end{figure*}

\begin{figure*}
	\includegraphics[width=0.99\textwidth]{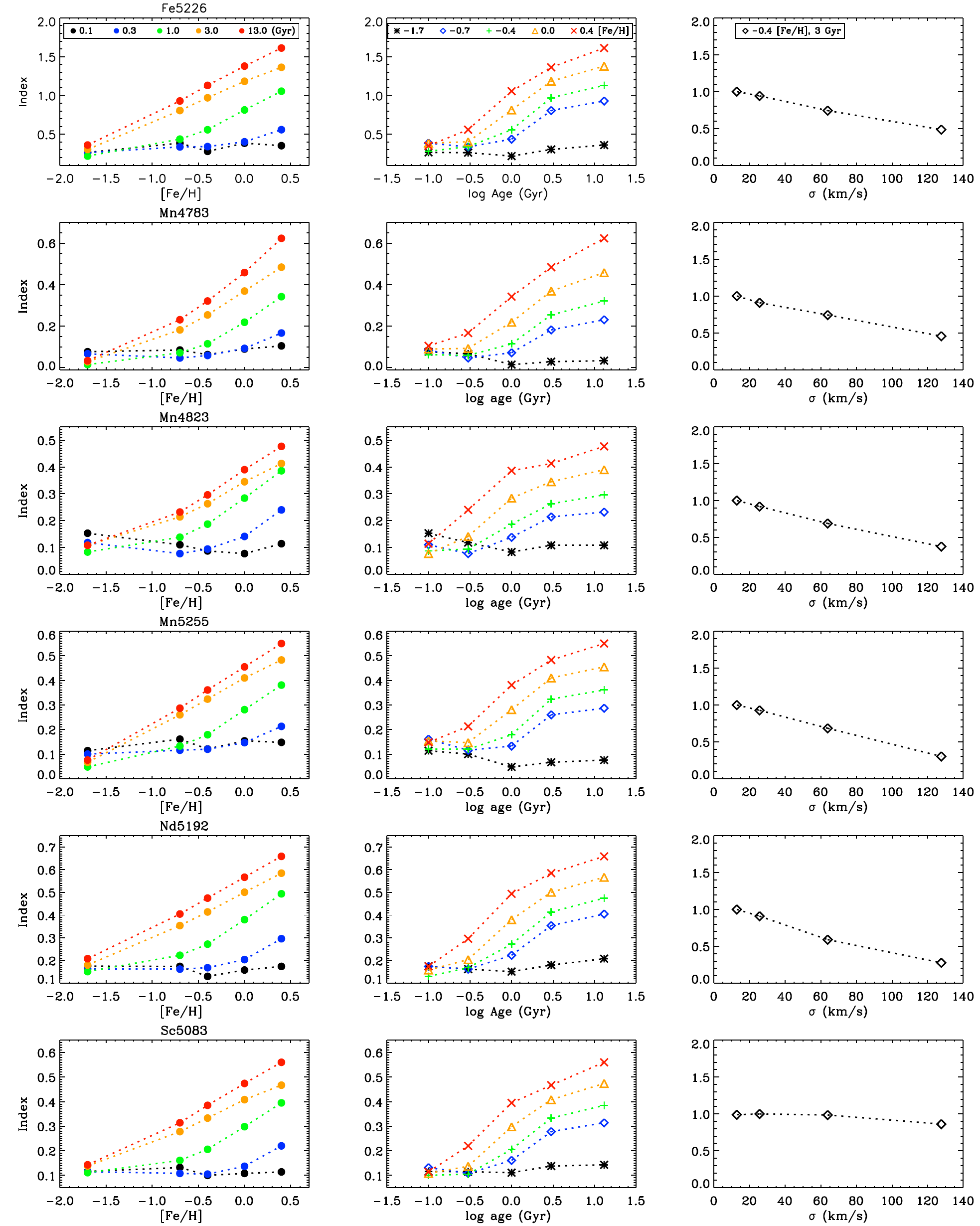}
    \caption{Same as Figure~\ref{fig:final_subfig_v3_2022}, now for the indices: Fe5226, Mn4783, Mn4823, Mn5255, Nd5192 and Sc5083.}
    \label{fig:final_subfig_v3_2_2022}
\end{figure*}

\begin{figure*}
	\includegraphics[width=0.99\textwidth]{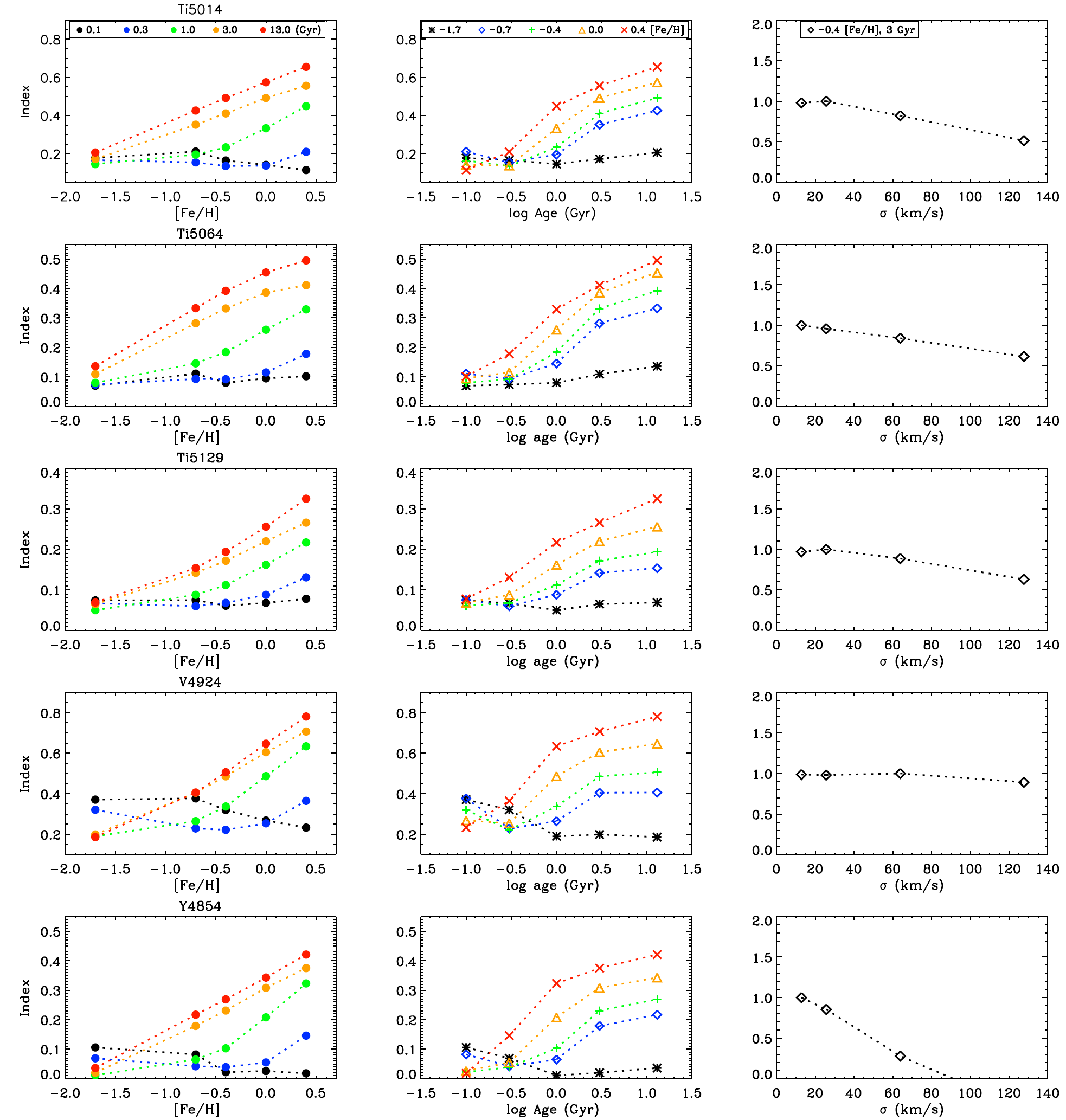}
    \caption{Same as Figure~\ref{fig:final_subfig_v3_2022}, now for the indices: Ti5014, Ti5064, Ti5129, V4924 and Y4854.}
    \label{fig:final_subfig_v3_3_2022}
\end{figure*}

\newpage
\section{All figures for the behaviour of the indices as a function of effective temperature, surface gravity and metallicity}
\label{sec:apC}
\begin{figure*}
	\includegraphics[width=0.98\textwidth]{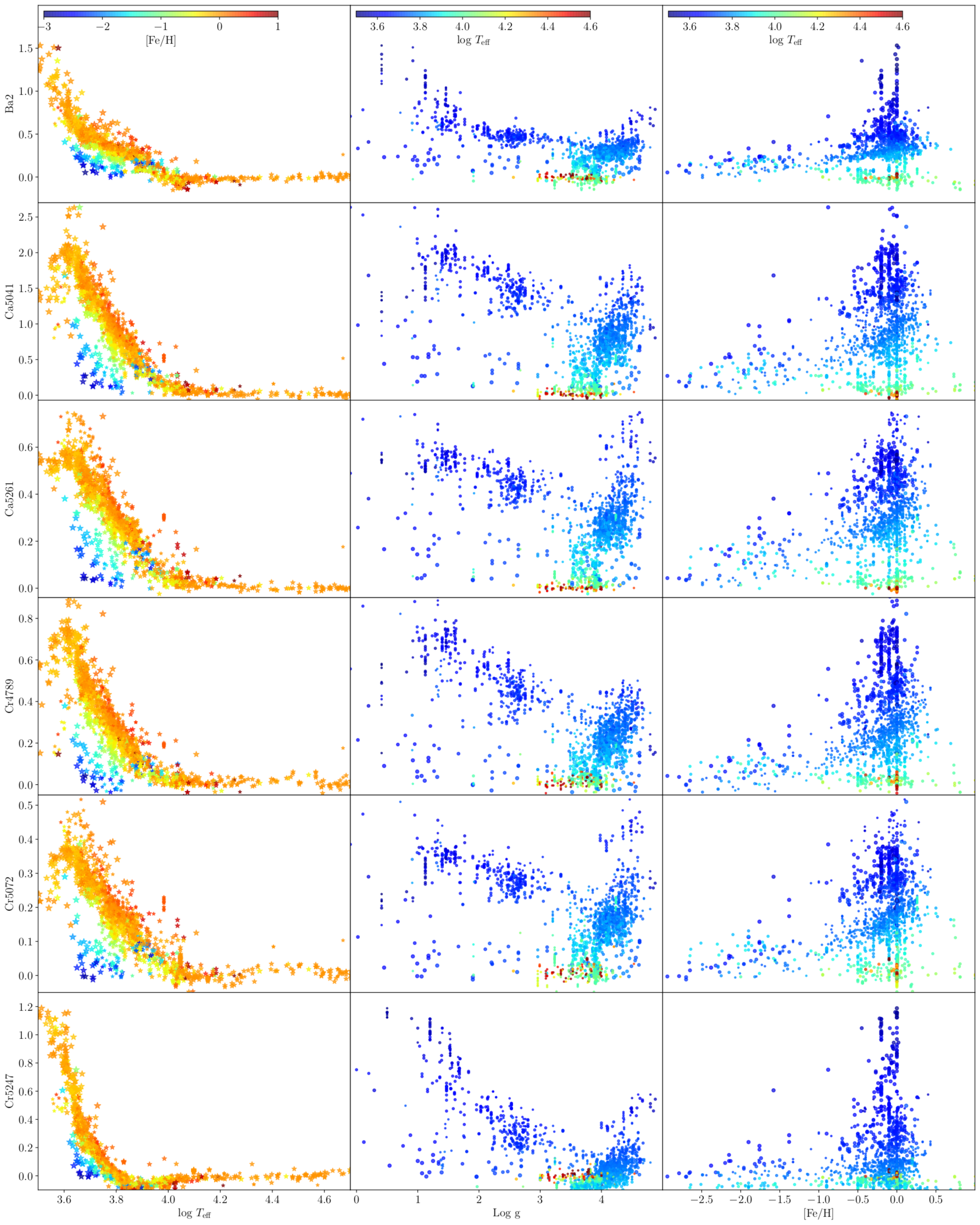}
    \caption{Indices as a function of effective temperature (log $T_\text{eff}$), surface gravity (log $\textit{g}$) and metallicity ([Fe/H]) shown for all the Elodie stars. The ELODIE library is a stellar database of 1959 spectra for 1503 stars with atmospheric parameters: $T_\text{eff}$ from 3000 K to 60000 K, log $\textit{g}$ from -0.3 to 5.9 and [Fe/H] from -3.2 to +1.4.}
    \label{fig:3d_elodie_forpaper}
\end{figure*}

\begin{figure*}
	\includegraphics[width=0.98\textwidth]{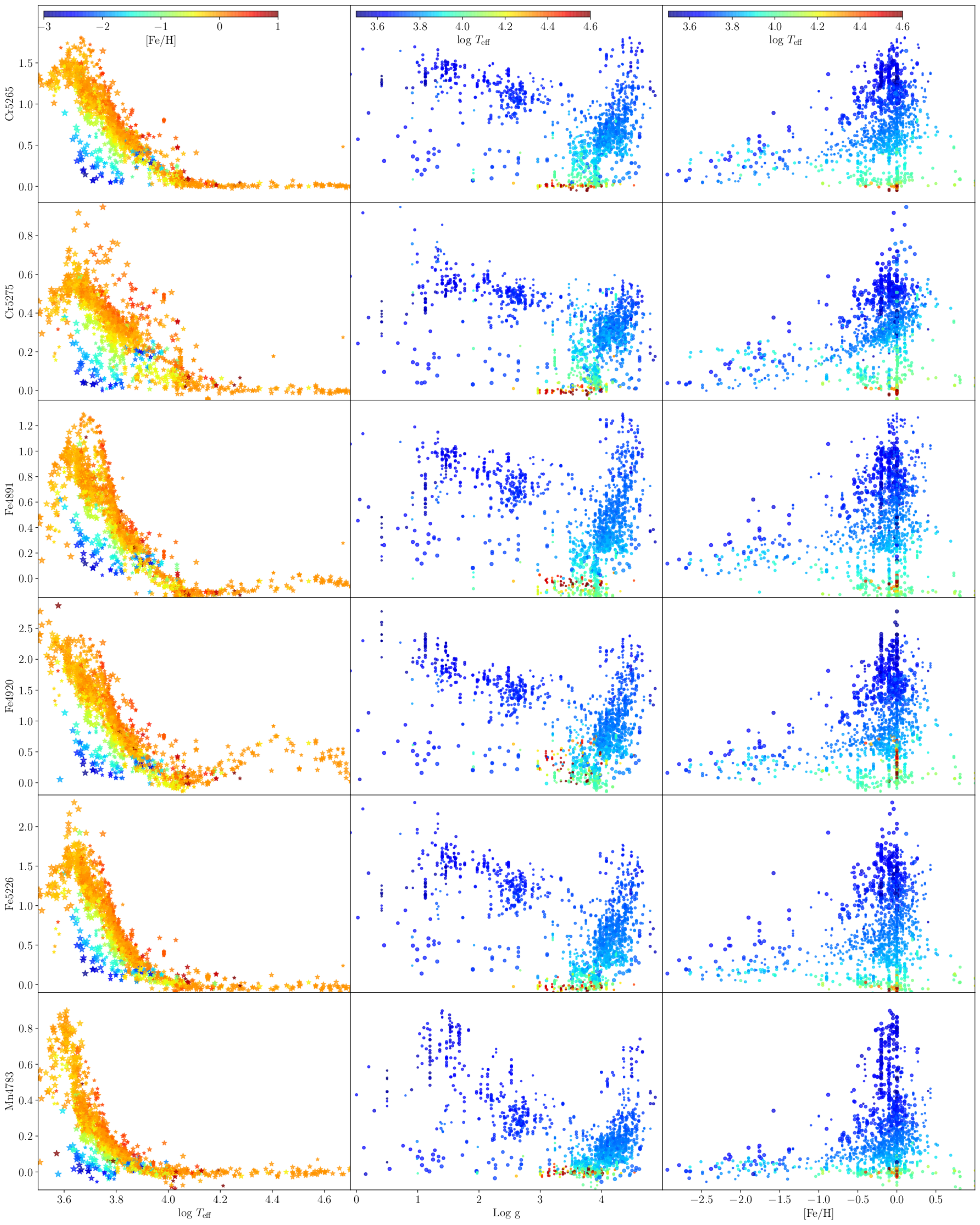}
    \caption{Same as in Figure~\ref{fig:3d_elodie_forpaper} but for Cr5265,  Cr5275, Fe4891, Fe4920, Fe5226 and Mn4783.}
    \label{fig:3d_elodie_2}
\end{figure*}

\begin{figure*}
	\includegraphics[width=0.98\textwidth]{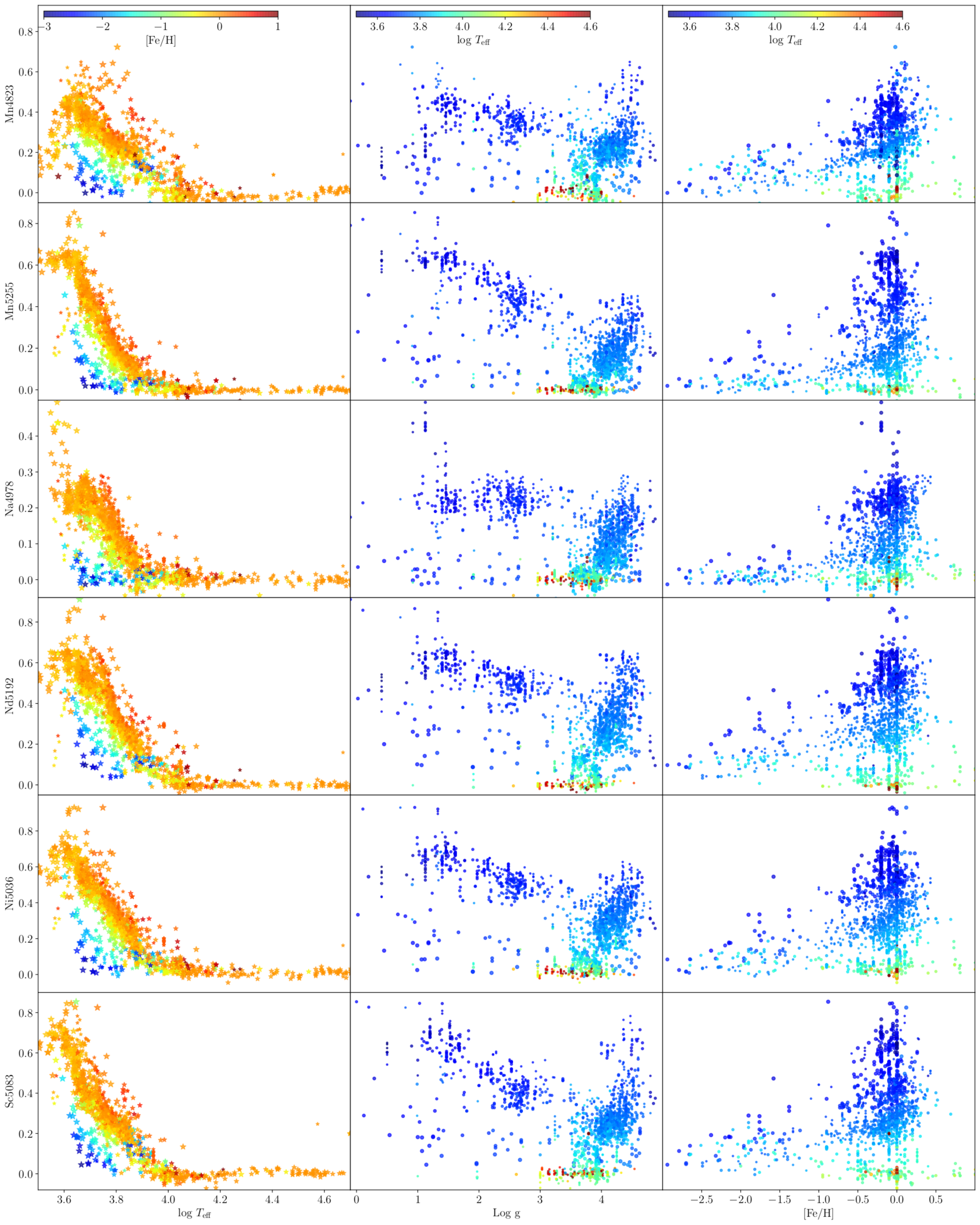}
    \caption{Same as in Figure~\ref{fig:3d_elodie_forpaper} but for Mn4823, Mn5255, Na4978, Nd5192, Ni5036 and Sc5083. }
    \label{fig:3d_elodie_3}
\end{figure*}

\begin{figure*}
	\includegraphics[width=0.98\textwidth]{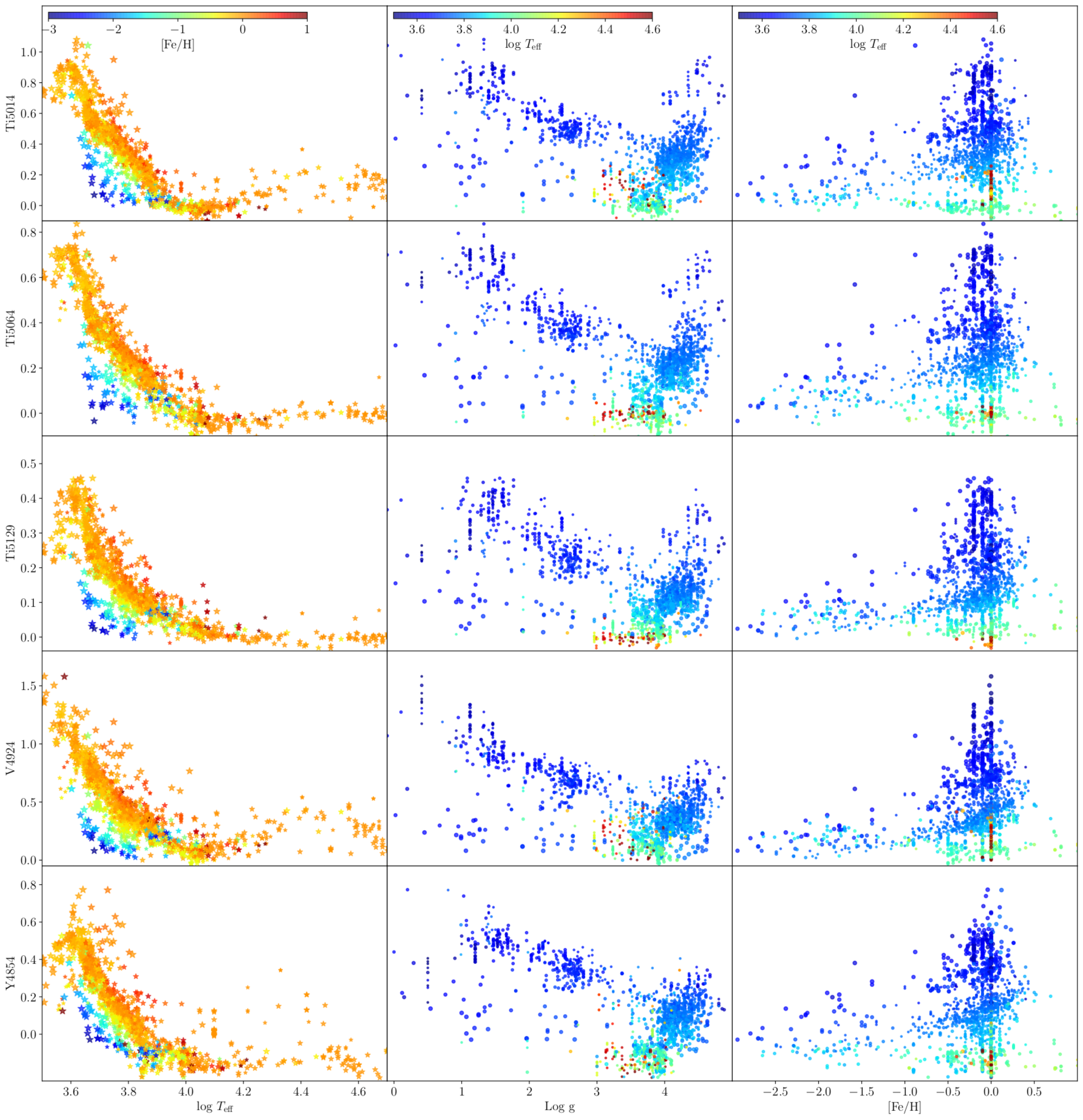}
    \caption{Same as in Figure~\ref{fig:3d_elodie_forpaper} but for Ti5014, Ti5064, Ti5129, V4924 and Y4854.}
    \label{fig:3d_elodie_4}
\end{figure*}

We measured the indices of Elodie stars with  $\sigma$= 25 kms$^{-1}$ to understand how the line indices change as a function of effective temperature (log $T_\text{eff}$), surface gravity (log $\textit{g}$) and metallicity ([Fe/H]). They are shown in Fig.~\ref{fig:3d_elodie_forpaper}-Fig.~\ref{fig:3d_elodie_4}.


\section{Figure for the comparison of age and metallicity dependence for elements}

\begin{figure*}
	\includegraphics[width=0.98\textwidth]{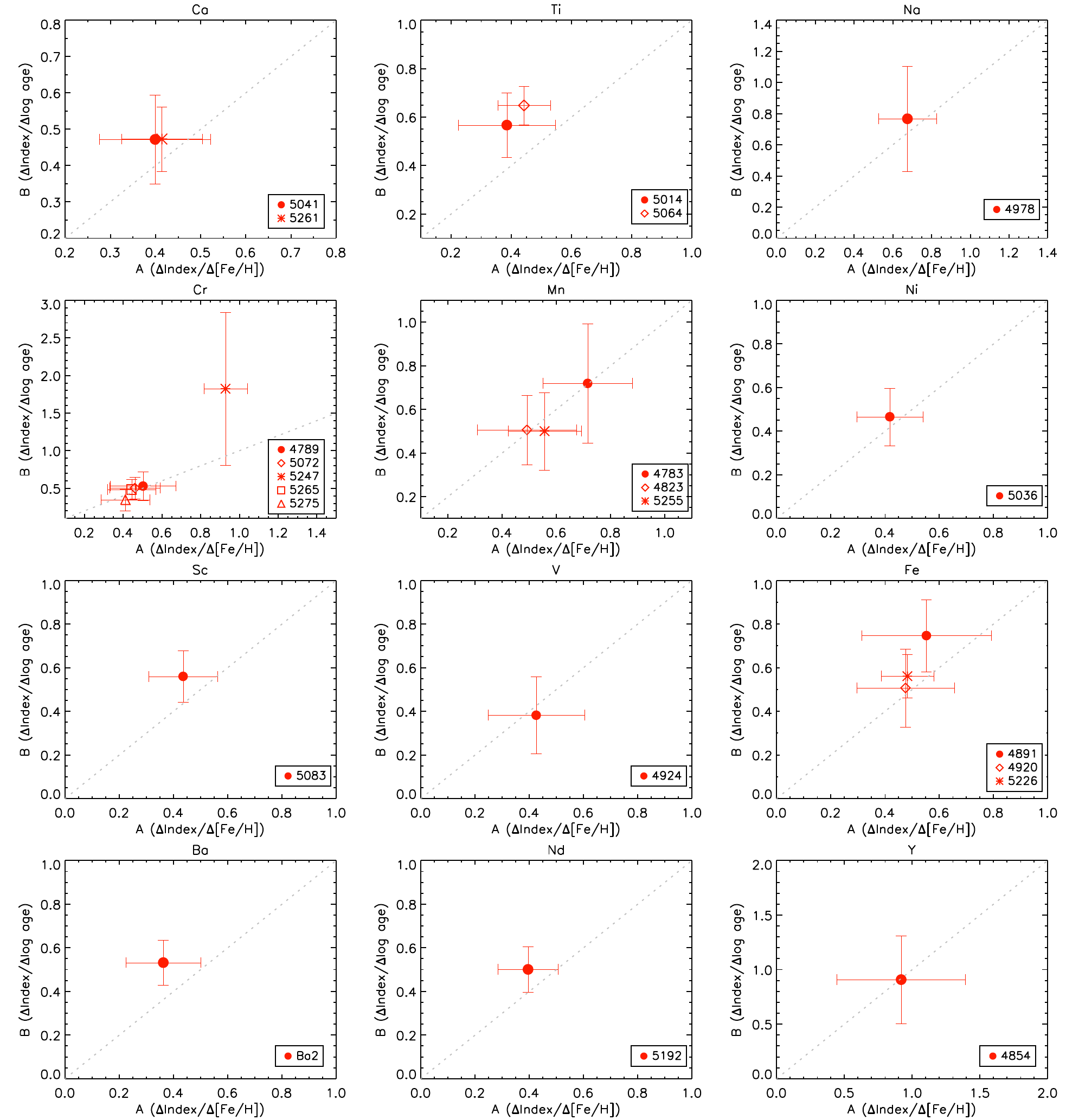}
    \caption{Comparison of age and metallicity dependence for Ca, Na, Y, Ba, Nd, Ti, Co, Cr, Mn, Ni, Sc, V and Fe elements. Data points near the X-axis are metallicity-indicators, while the ones near the Y-axis are age indicators.}
    \label{fig:final_a_and_b_elements_alpha}
\end{figure*}

Figure~\ref{fig:final_a_and_b_elements_alpha} shows the values of A${_i}$ and B${_i}$ for each index ${i}$ grouped per element.

\section{Comparison models with MILES and PEGASE}
We compare the line indices measured from the PEGASE and MILES SSPs.
\begin{figure*}
	\includegraphics[width=0.98\textwidth]{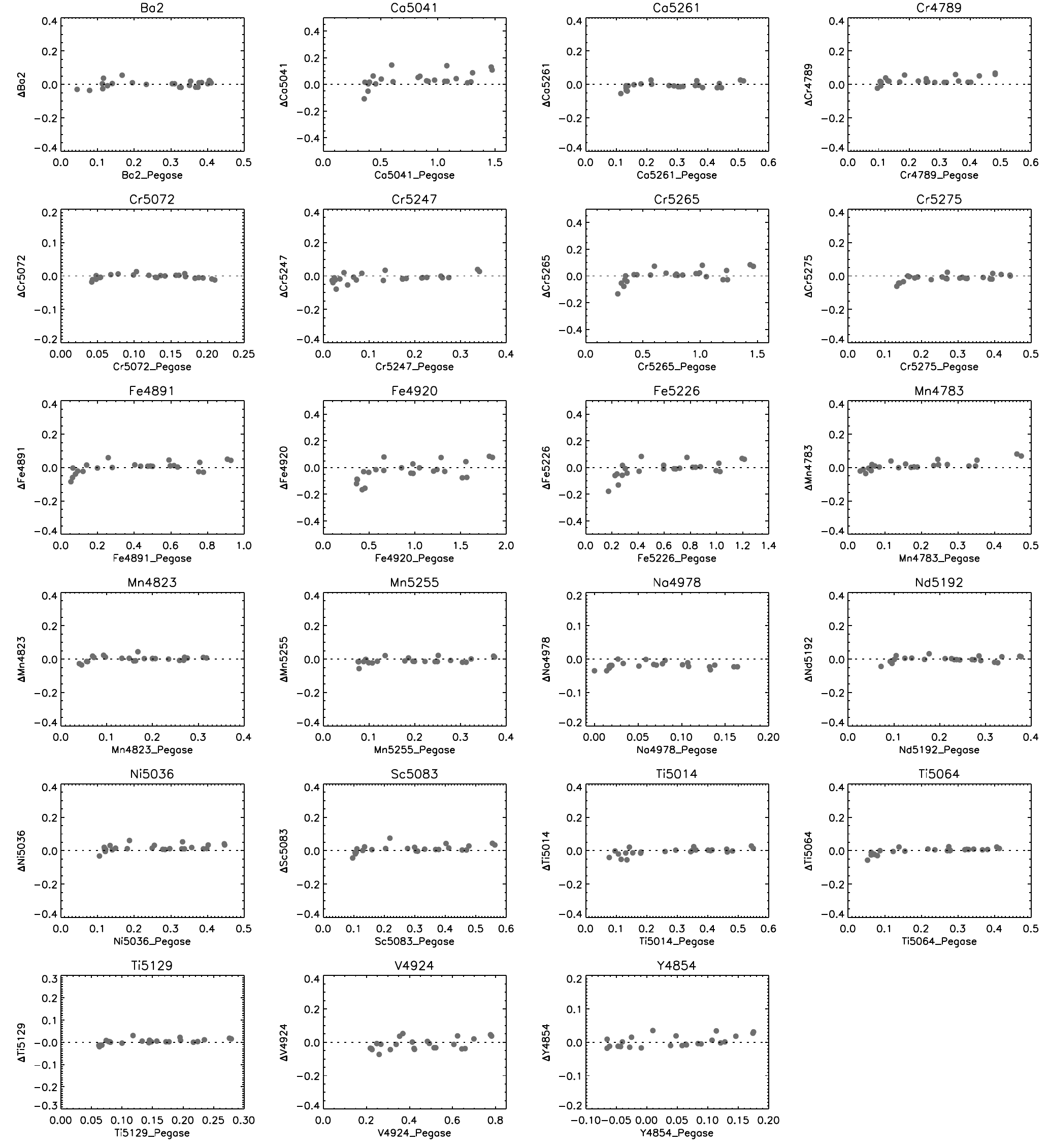}
    \caption{Our indices measured from MILES and PEGASE with 60km/s Here, $\Delta{index}$ =  $index_\text{PEGASE}$ - $index_\text{MILES}$}.
    \label{fig:comp_Miles_Pegase}
\end{figure*}


\bsp	
\label{lastpage}
\end{document}